\documentclass[twocolumn]{autart}    

\usepackage{graphicx}          
\usepackage{amsmath} 
\usepackage{amsfonts}  
\usepackage{derivative}  
\usepackage{mathtools} 
\usepackage{arydshln}  
\usepackage{xcolor} 
\usepackage{lscape} 
\usepackage{bm}  
\usepackage[utf8]{inputenc}
\usepackage{mathtools}
\usepackage{algorithm}
\usepackage{algpseudocode}

\date{25 July 2025}

\definecolor{JLcolor}{rgb}{1,0.3,0}

\begin{document}

\begin{frontmatter}

\title{Kernel-Based Sparse Additive Nonlinear Model Structure Detection through a Linearization Approach  \thanksref{footnoteinfo}} 

\thanks[footnoteinfo]{This paper was not presented at any IFAC 
meeting. 
This research was supported by the Fund for Scientific Research (FWO Vlaanderen, grant G005218N) and by the Flemish Government (Methusalem Grant METH1).
This paper was not presented at any IFAC meeting. Corresponding author Sadegh Ebrahimkhani.}

\author[VUB]{Sadegh Ebrahimkhani}\ead{Sadeghebrahimkhani@gmail.com},               
\author[VUB]{John Lataire}\ead{John.Lataire@vub.be}  

\address[VUB]{dept.\ ELEC, Vrije Universiteit Brussel, Pleinlaan 2, 1050, Brussels, Belgium}


\begin{abstract}                          
The choice of parameterization in Nonlinear (NL) system models greatly affects the quality of the estimated model. Overly complex models can be impractical and hard to interpret, necessitating data-driven methods for simpler and more accurate representations. In this paper, we propose a data-driven approach to simplify a class of continuous-time NL system models using linear approximations around varying operating points. Specifically, for sparse additive NL models, our method identifies the number of NL subterms and their corresponding input spaces. Under small-signal operation, we approximate the unknown NL system as a trajectory-scheduled Linear Parameter-Varying (LPV) system, with LPV coefficients representing the gradient of the NL function and indicating input sensitivity. Using this sensitivity measure, we determine the NL system’s structure through LPV model reduction by identifying non-zero LPV coefficients and selecting scheduling parameters. We introduce two sparse estimators within a vector-valued Reproducing Kernel Hilbert Space (RKHS) framework to estimate the LPV coefficients while preserving their structural relationships. The structure of the sparse additive NL model is then determined by detecting non-zero elements in the gradient vector (LPV coefficients) and the Hessian matrix (Jacobian of the LPV coefficients). We propose two computationally tractable RKHS-based estimators for this purpose. The sparsified Hessian matrix reveals the NL model’s structure, with numerical simulations confirming the approach’s effectiveness.
\end{abstract}

\end{frontmatter}

\section{Introduction}
Nonlinear (NL) modeling provides greater flexibility and accuracy when Linear Time-Invariant (LTI) approximations fail to capture dynamic system behavior. NL system identification plays a crucial role in analysis, modeling, and control design across diverse scientific domains \cite{van2020data,schoukens2019nonlinear}. However, estimating accurate yet simple NL models remains challenging in many scientific and engineering fields \cite{zhou2021gene,ndaoud2020optimal}. The choice of NL model parametrization greatly influences model quality. Identifying the key input variables that most affect the NL function describing system dynamics is vital, as is determining the number of NL subterms and their dependencies on inputs to ensure accurate, simple, and interpretable models.

The key challenge in structure detection is choosing a subset of candidate terms that both accurately estimates  the output and keeps the model compact. Existing approaches include sparse‐regression–based discovery using predefined function libraries \cite{brunton2016discovering,lai2019sparse}, AIC (Akaike Information Criterion)‐driven selection \cite{mangan2017model}, and polynomial NARX (Nonlinear AutoRegressive eXogenous Input) model reduction via decoupling of polynomial functions  \cite{dreesen2015decoupling,westwick2018using} or pruning \cite{billings2013nonlinear}. Methods such as LASSO (Least Absolute Shrinkage and Selection Operator) \cite{kukreja2006least,tibshirani1996regression}, Non‐Negative Garrote (NNG) \cite{breiman1995better}, and SPARSEVA \cite{rojas2014sparse} add penalties (often an $\ell_1$ norm) during estimation, but their performance hinges on the initial choice of the library or set of basis functions.

In this article, we address structure detection for continuous-time NL systems via small-signal linearization around varying operating points~\cite{ebrahimkhani2025combined,Allen3,JointPaper,CBs1,van2018revealing,DSR1}. This divide-and-conquer strategy~\cite{JointPaper,DandC}--previously used for block-oriented model structure discrimination~\cite{schoukens2015structure} and NL model reduction~\cite{ebrahimkhani2025continuous}--breaks NL model identification into manageable subproblems solvable via dedicated    tools~\cite{toth2010modeling,LPVBFE}. The NL system’s response to input perturbations around a varying operating point can be approximated by a trajectory‑scheduled LPV model~\cite{ebrahimkhani2025combined,Sadeghifac,CBs1,JointPaper,Allen3}, whose coefficient vector equals the gradient of the unknown NL function \cite{ebrahimkhani2025combined,JointPaper}. This gradient relationship defines structural constraints on the LPV coefficients, which we exploit for NL model structure detection. Using a sensitivity measure, we propose two estimators for LPV model reduction: first, linking LPV model order selection (identifying nonzero coefficients) to input-variable selection for the NL system; second, performing LPV scheduling selection by identifying the nonzero elements of a Hessian matrix.

The reduction of LPV models involves two main challenges: choosing the model order and reducing the dimension of the scheduling variables \cite{rizvi2016kernel}. Model‐order selection often relies on sparse estimation—typically via an $\ell_1$‐norm penalty—to identify a suitable LPV structure \cite{piga2013lpv}. Methods such as NNG \cite{breiman1995better,toth2009order}, LASSO \cite{tibshirani1996regression}, and SPARSEVA \cite{rojas2014sparse} promote sparsity in the estimated coefficients, though their success depends on a rich basis‐function parametrization. 
Reducing the scheduling dimension has been addressed by techniques that project inputs onto a lower‐dimensional subspace \cite{olucha2024reduction}. Common approaches include Principal Component Analysis (PCA) \cite{kwiatkowski2008pca}, Kernel PCA \cite{rizvi2016kernel}, autoencoders \cite{rizvi2018model}, and deep neural networks \cite{koelewijn2020scheduling}, all of which seek compact yet expressive representations.

More generally, variable selection in linear or additive models is well established \cite{hastie2015statistical,koltchinskii2010sparsity}, but inducing sparsity in NL functions remains challenging \cite{yamada2014high,chen2017kernel}. Gradient‐based input‐variable selection has proven effective in discrete‐time NL systems \cite{bai2019variable,cheng2021variable}, in NARX models \cite{ashraf2024partial}, and for neural‐network pruning \cite{engelbrecht2001new}.

RKHS offers a flexible nonparametric framework for function estimation, balancing bias–variance trade‐offs, enhancing robustness, and aiding in model‐order selection \cite{RKHSgnr2}. It also enables seamless integration of prior knowledge into the model \cite{10990161,khosravi2023kernel}. In this paper, we formulate LPV model‐reduction estimators within an RKHS of gradient functions.

The discrete‐time LPV model–order‐selection problem in an RKHS has been studied in \cite{laurain2020sparse}, where coefficient sparsity is formulated as a regularized estimator. We extend this to continuous‐time systems using a vector‐valued RKHS of gradient vector fields, thereby enforcing interdependence among LPV coefficients and enabling sensitivity‐based input‐variable selection for the NL system. Additionally, we adopt the approach of \cite{rosasco2013nonparametric} by introducing a regularized sparse estimator to reduce the dimension of scheduling variables in the LPV model. This estimator uses partial‐derivative regularization to promote sparsity. We extend the method to a vector‐valued RKHS of gradient vector fields, showing that LPV‐scheduling selection can be reformulated as identifying the non‐zero elements of a Hessian matrix. Exploiting the structural symmetry of this matrix, we reduce the number of regularization terms. The resulting sparsified Hessian is then used for NL model‐structure detection.

We consider the direct identification of continuous-time LPV systems within the frequency domain. By employing a mixed time-frequency domain formulation \cite{LTVRKHS,LTVBFE,LPVBFE}, we take advantage of the flexibility in selecting the frequency band of interest and the precise computation of time derivatives.

In this paper, we present the following contributions:
\begin{enumerate}
\item Structure detection for a class of sparse additive continuous-time NL system models.

\item  Formulation of both input variable selection for the NL system and scheduling selection for its LPV approximation as two tractable convex optimization problems.
\end{enumerate}

The paper is organized as follows. In Section~\ref{sec:ProbFor}, we state the key assumptions and outline the problem setting. Section~\ref{sec:FreqDom} introduces the frequency–domain model of the LPV system, and Section~\ref{sec:LPVparam} presents the LPV parameterization. Basic concepts of RKHS are reviewed in Section~\ref{sec:vectorVRKHS}. Section~\ref{Sec:PropMeth} describes our proposed model-reduction method for the LPV system, and Section~\ref{Sec:NLmdlreduced} establishes the connection between the reduced LPV model and the NL system’s structure. Section~\ref{sec:Simres} demonstrates the proposed approach through simulations, and Section~\ref{sec:conclu} provides concluding remarks.

\textbf{Notation}
Let $\mathbb{R}$, $\mathbb{N}$ and $\mathbb{C}$ denote the sets of real numbers, positive integers and complex numbers. For $n\in\mathbb{N}$, the standard inner product on $\mathbb{R}^n$ is denoted by $\langle x,y\rangle$, and $\|x\|_{p}$ the $\ell_{p}$–norm. The space of $n$–times continuously differentiable functions is $\mathcal{C}^n$. Superscripts $(n)$, $\top$ and $H$ denote the $n$-th time‐derivative, transpose and Hermitian transpose, respectively. Bold upper‐ (resp.\ lower‐) case letters denote matrices (resp.\ vectors); normal letters denote scalars. Floor and ceiling functions are $\lfloor\cdot\rfloor$ and $\lceil\cdot\rceil$. The $n\times n$ identity matrix is $\mathbf{I}_{n}$, $\odot$ the Hadamard product, and $\mathbb{N}_{n} = \{1,\dots,n\}$. For $\mathbf{f}:[0,\infty)\to\mathbb{R}^n$,  $\|\mathbf f\|_{\mathcal L_\infty}
=\sup_{t\ge0}\|\mathbf f(t)\|$,  and $\mathcal L_\infty^n$ is the set of smooth $\mathbf f$ whose truncation   $\mathbf f_\tau(t) =
\begin{cases}
\mathbf f(t) & 0 \le t \le \tau\\
0            & t > \tau
\end{cases}$, with $\tau\in[0,\infty)$ satisfies $\|\mathbf f\|_{\mathcal L_\infty}<\infty$ \cite[Section 5.1]{KhalilB}.
 A continuous function $f:[0,\tau)\to[0,\infty)$ belongs to class $\mathcal{K}$ if it is strictly increasing with $f(0)=0$ \cite[Definition 4.2]{KhalilB}. It belongs to class $\mathcal{KL}$ if, for each fixed $t$, $f(\cdot,t)\in\mathcal{K}$, and for each fixed $s$, $f(s,t)$ decreases to~0 as $t\to\infty$ \cite[Definition 4.3]{KhalilB}.
We adopt the following conventions for the notation `$\mathrm{diag}$': 
$\mathrm{diag}(\mathbf{x})$: diagonal matrix with vector $\mathbf{x}$ on the diagonal.
$\mathrm{diag}(\mathbf{X})$: vector of diagonal entries of matrix $\mathbf{X}$.
$\mathrm{diag}(\mathbf{X}_1, \dots, \mathbf{X}_n)$: block-diagonal matrix with blocks $\mathbf{X}_1, \dots, \mathbf{X}_n$.

\section{Problem Formulation} \label{sec:ProbFor}
We consider a class of continuous-time NL systems described by the following NL Ordinary Differential Equation (ODE):
\begin{equation} \label{eq:NLsys}
 y^{(n_a+1)}(t)=f\left(y(t), \dots, y^{(n_a)}(t), u(t), \dots, u^{(n_b)}(t)\right) 
\end{equation}
where $u(t)$ and $y(t)$ denote the system’s input and output at time $t$, respectively. The unknown function $f: \mathcal{U} \rightarrow \mathbb{R}$, with $\mathcal{U} \subseteq \mathbb{R}^{n_x}$ where $n_x=n_a+n_b+2$.
\begin{defn}[Varying operating point] \label{defn:TimeVarOpPoint}
We define the `large input' signal $u_L(t)$, which drives the NL system \eqref{eq:NLsys} to a varying operating point (large trajectory):
\begin{equation} \label{eq:PL}
\mathbf{p}_L(t) \coloneqq \begin{bmatrix} y_L(t) & \dots & y_L^{(n_a)}(t) & u_L(t) & \dots & u_L^{(n_b)}(t) \end{bmatrix} ^{\top}
\end{equation}
\end{defn}
The selection of the operating point in Definition~\ref{defn:TimeVarOpPoint} should be designed or chosen based on the specific application, for instance, to encompass the envelope of the operational range of the underlying system. 
\begin{assum}  \label{assum:SparsityofNLf}
For the NL system in \eqref{eq:NLsys} and the operational range defined in Definition~\ref{defn:TimeVarOpPoint}, the NL function can be decomposed into a sum of subfunctions:
\begin{equation} \label{eq:basicLowcomplxfunc}
f\left(y(t), \dots, y^{(n_a)}(t), u(t), \dots, u^{(n_b)}(t)\right)=\sum_{i=1}^{M} f_{i}(\mathbf{p}_i)
\end{equation}
with $f_i : \mathcal{U}_i \rightarrow \mathbb{R}$, $\mathbf{p}_i \in \mathcal{U}_i \subset \mathcal{U}$, where $\mathcal{U}_i \cap \mathcal{U}_j = \emptyset$ for $i \neq j$ and $i, j \in \mathbb{N}_M$, and $\bigcup_{i=1}^M \mathcal{U}_i \subseteq \mathcal{U}$. Here, $M > 1$ and $M \leq n_x$ (there are at least two subfunctions in \eqref{eq:basicLowcomplxfunc}), and no subfunction $f_i$ can be expressed as a combination of the others. The number of subfunctions, $M$, and their input spaces, $\mathcal{U}_i$, are unknown. 
\end{assum}
\begin{assum} \label{assum:simplyconnected}
The domain $\mathcal{U}$ is simply connected.
\end{assum}
\begin{rem}
A domain $\mathcal{U}$ is simply connected if every loop in $\mathcal{U}$ can be continuously contracted to a point within $\mathcal{U}$.  For non‐closed trajectories, one may choose $\mathcal{U}$ as a tubular neighborhood of the trajectory.  For closed trajectories (limit cycles), $\mathcal{U}$ is taken as a neighborhood enclosing the trajectory, which we assume to be simply connected.
\end{rem}

\begin{assum} \label{assum:fisinC2clss}
The unknown function $f \in \mathcal{C}^2 $ locally  on  $ \mathcal{U} $, with all its first- and second-order partial derivatives bounded.
\end{assum}

\begin{defn}[Input-to-Output Stability \cite{sontag1999notions}]
The NL system \eqref{eq:NLsys} is said to be  Input-to-Output (IO) stable, if there exist finite constant $ \eta_0 > 0 $,  $ \eta_1 \in \mathcal{K} $, and $ \eta_2 \in \mathcal{KL} $  such that:
\begin{equation*} \label{eq:LinfsmallsigSt11}
\begin{aligned}
& \sup_{ 0  \leq t  \leq  a}  \|u(t)\| \leq \eta_0, a \in [0,\infty), \quad  u \in \mathcal{L}_{\infty}^1 \implies y \in \mathcal{L}_{\infty}^1, \; \\
&{\scriptsize
  \|y\|_{\mathcal{L}_{\infty}} \leq \eta_1 ( \|u\|_{\mathcal{L}_{\infty}}) + \eta_2\Big(\|[y(0), y^{(1)}(0), \ldots, y^{(n_a)}(0)]\|, t\Big) }
\end{aligned}
\end{equation*}
Here, the norm of the input in $\sup$ can be any $p$-norm.
\end{defn}

\begin{assum} \label{assum:BIBOstable}
The NL system \eqref{eq:NLsys} is locally IO stable around  the operating point $\mathbf{p}_L(t)$.
\end{assum}
The IO stability of \eqref{eq:NLsys} ensures that bounded inputs produce bounded outputs. Moreover, since $\eta_{1}(\cdot)$ is strictly increasing, choosing the input `sufficiently' small in the $\mathcal{L}_{\infty}$‐norm guarantees a correspondingly small output. 

\begin{defn}[Class of small input signals $S_{\varepsilon}$] \label{defn:SmallInsignal}
We define $S_{\varepsilon} \subset \mathcal{L}^1_\infty$ as the class of small input perturbations, and a signal $\tilde{u}(t) \in S_{\varepsilon}$ if:
\begin{equation}
 \| \tilde{u}_{\tau}(t) \|_{\mathcal{L}_{\infty}} = \varepsilon, \quad \varepsilon > 0
\end{equation}
where $\tau \in [0, \infty)$ denotes the measurement time, and $\tilde{u}_{\tau}(t)$ is defined as the truncation of $\tilde{u}(t)$ over the interval $[0, \tau]$, i.e.,
$
\tilde{u}_{\tau}(t) =
\begin{cases}
\tilde{u}(t)  &   0 \le t \le \tau  \\
0  &  t > \tau
\end{cases}
$.
\end{defn}
Under Assumption~\ref{assum:BIBOstable}, the NL system's trajectory stays near $\mathbf{p}_L(t)$ under small input perturbations $\tilde{u}(t) \in S_{\varepsilon}$ (for sufficiently small $\varepsilon$). Thus, the system's input and output are:
\begin{subequations}
\begin{align}
u(t) &= u_L(t) + \tilde{u}(t) \label{eq:largeplussmallIN} \\
y(t) &= y_L(t) + \tilde{y}(t) \label{eq:largeplussmallOOUT}
\end{align}
\end{subequations}
where $\tilde{y}(t)$ is the `small output' signal. 
Given the large trajectory from Definition~\ref{defn:TimeVarOpPoint} and its perturbed form $\mathbf{p}(t) = \mathbf{p}_L(t) + \mathbf{\tilde{p}}(t)$, where $\mathbf{\tilde{p}}(t)$ is the small-signal vector:
\begin{equation} \label{eq:SmallTraj}
\mathbf{\tilde{p}}(t) \coloneqq \begin{bmatrix} \tilde{y}(t) & \dots & \tilde{y}^{(n_a)}(t) & \tilde{u}(t) & \dots & \tilde{u}^{(n_b)}(t) \end{bmatrix}^{\top}
\end{equation}
The NL system \eqref{eq:NLsys} around $\mathbf{p}_L(t)$ can be rewritten using Taylor's expansion theorem (see \cite{DSR1} and \cite[Section 4.9]{DSR2}) as:
\begin{equation} \label{eq:TaylorExpThm}
\begin{aligned}
& y^{(n_a+1)}(t) = y_L^{(n_a+1)}(t)+\tilde{y}^{(n_a+1)}(t)=f(\mathbf{p}(t)) \\
&= f(\mathbf{p}_L(t)) + ( \nabla f(\mathbf{p}_L(t)) )^\top \mathbf{\tilde{p}} (t) + \mathcal{E}(\mathbf{\tilde{p}}(t),t)
\end{aligned}
\end{equation}
This is an alternative form of the original NL system, with $\mathcal{E}(\mathbf{\tilde{p}}(t), t)$ representing the higher-order terms (details in \cite{DSR1} and \cite[Section 4.9]{DSR2}).

\begin{defn}[Linear approximation] \label{defn:Linearapprox}
Neglecting the remainder term $\mathcal{E}(\cdot)$ in \eqref{eq:TaylorExpThm} and considering $y_L^{(n_a+1)}(t)=f(\mathbf{p}_L(t))$, the linear approximation of the NL system \eqref{eq:NLsys} around $\mathbf{p}_L(t)$ is defined as:
\begin{equation} \label{eq:LPV1}
\begin{aligned}
& \tilde{y}^{(n_a+1)}(t) \approx \tilde{y}_0^{(n_a+1)}(t) =  (\nabla f(\mathbf{p}_L))^\top \mathbf{\tilde{p}}_0 (t)= \\ 
& \sum^{n_a}_{n=0} a_{n}(\mathbf{p}_L(t)) \frac{d^{n} \tilde{y}_0(t)}{dt^{n}}+ 
 \sum^{n_b}_{m=0} b_{m}(\mathbf{p}_L(t)) \frac{d^{m}\tilde{u}(t)}{dt^{m}}
\end{aligned}
\end{equation}
where:
\begin{equation}
\mathbf{\tilde{p}}_0(t) \coloneqq \begin{bmatrix} \tilde{y}_0(t) & \dots & \tilde{y}_0^{(n_a)}(t) & \tilde{u}(t) & \dots & \tilde{u}^{(n_b)}(t) \end{bmatrix}^\top
\end{equation}
with $\tilde{y}_0(t)$ represents the output of the linear approximation system, and the LPV coefficients are:
\begin{equation} \label{eq:PV1}
 a_{n}(\mathbf{p}_L(t))= \left. \frac{\partial f}{\partial y^{(n)}}\right |_{\mathbf{p}_L(t)}, \,
 b_{m}(\mathbf{p}_L(t))=\left. \frac{\partial f}{\partial u^{(m)}}\right |_{\mathbf{p}_L(t)}
\end{equation}
\end{defn}
We define the vector of LPV coefficients as
\begin{equation} \label{eq:PV2}
\begin{aligned}
& \mathbf{c}(\mathbf{p}_L) \coloneqq \begin{bmatrix}
c_{1}(\mathbf{p}_L) & c_{2}(\mathbf{p}_L) & \dots & c_{n_x}(\mathbf{p}_L)
\end{bmatrix} ^\top = \\
& {\scriptsize
\begin{bmatrix}
a_{0}(\mathbf{p}_L) & a_{1}(\mathbf{p}_L) & \dots & a_{n_a}(\mathbf{p}_L) & b_{0}(\mathbf{p}_L) & b_{1}(\mathbf{p}_L) & \dots & b_{n_b}(\mathbf{p}_L)
\end{bmatrix}} ^\top
\end{aligned}
\end{equation}

\begin{assum} \label{assum:BIBOLPVstbl}
The linear  system \eqref{eq:LPV1} is locally bounded input bounded output  stable around the origin.
\end{assum}


\begin{thm} [\cite{DSR1} and Section 4.9 in \cite{DSR2}] \label{thm:linearapprox}
Under the Assumptions~\ref{assum:fisinC2clss}, \ref{assum:BIBOstable}, and \ref{assum:BIBOLPVstbl}, when $\varepsilon$ is sufficiently small ($\tilde{u}(t) \in S_\varepsilon$), the peak absolute difference between the states of the linear approximation \eqref{eq:LPV1} and those of the NL system \eqref{eq:TaylorExpThm} which is defined as:
\begin{equation} \label{eq:apperr}
\kappa \coloneqq  \left\lVert
\begin{bmatrix} \tilde{y}(t) & \dots & \tilde{y}^{(n_a)}(t) \end{bmatrix}
-
\begin{bmatrix} \tilde{y}_0(t) & \dots & \tilde{y}_0^{(n_a)}(t) \end{bmatrix} \right\rVert_{\mathcal{L}_{\infty}}
\end{equation}
 remains bounded. Furthermore, $\kappa$ can be minimized to an arbitrarily small value by selecting a sufficiently small $\varepsilon$. In addition, $\tilde{y}(t) = \tilde{y}_0(t) + \mathcal{O}(\varepsilon^2)$.
\end{thm}




\begin{rem}
The choice of perturbation signal~$\tilde u(t)$ depends on the system under study.  Its amplitude may be set experimentally or via manual tuning \cite{LTPfNL,van2018revealing,hess2014respiratory,LTVfNL}, or treated as a design parameter \cite{CBs2}.  Typically, $\tilde u(t)$ is broadband: when the operating point changes slowly or fails to excite the desired frequencies, a broadband perturbation is superimposed on the nominal input to probe the linear dynamics over a wide band.  Applications include time‐varying mechanical impedance measurement \cite{van2018revealing}, electrochemical impedance spectroscopy \cite{lee2024small,chang2010electrochemical}, grid impedance measurement \cite{8194844}, respiratory oscillometry \cite{keymolen2023low}, and power converter analysis \cite{corradini2015digital}.
 For a detailed analysis of stability requirements, and small-signal operation, please refer to \cite{ebrahimkhani2025combined,DSR1}.
\end{rem}


Note that the LPV model \eqref{eq:PV1} admits an LTV representation with coefficients $c_j(t) = c_j(\mathbf{p}_L(t))$ evaluated at each $t$. We denote LTV coefficients as $c_j(t)$ and LPV coefficients as $c_j(\mathbf{p}_L(t))$, with domains $\mathbb{R}_{\ge 0}$ and $\mathcal{U}$, respectively. We assume the approximation error in \eqref{eq:apperr} is sufficiently small—by constraining the experiment—so that $\tilde y_0 \approx \tilde y$ .Unless otherwise noted, we omit the explicit time argument in $\mathbf p_L(t)$.

We simplify the NL model by reducing the LPV model~\eqref{eq:LPV1}.  In particular, we determine the number of subfunctions $M$ and their domains $\mathcal{U}_i$, $i\in\mathbb{N}_M$, according to Assumption~\ref{assum:SparsityofNLf}.  This involves two steps: choosing the model order in \eqref{eq:LPV1} and selecting the scheduling variables in the LPV coefficients $c_j(\mathbf p_L)$ for $j\in\mathbb{N}_{n_x}$.  Together, these steps yield a lower‐complexity LPV approximation, and realizing the sparse additive decomposition of Assumption~\ref{assum:SparsityofNLf}.
\section{Frequency Domain Model}\label{sec:FreqDom}
In this section, we derive the frequency-domain model of the continuous-time LPV system \eqref{eq:LPV1}. Our approach follows the methodology presented in \cite{lataire2011frequency,LTVRKHS,LPVBFE} to reformulate the continuous-time (LTV/LPV) ODE in the frequency domain.
\begin{defn}[Dataset] \label{defn:DataSet}
The estimation dataset comprises $N$ samples of the large trajectory and the corresponding small input and output:
\begin{equation} \label{eq:DataSet0}
\mathcal{D} \coloneqq \left\{ \left(\tilde{u}(t), \tilde{y}(t), \mathbf{p}_L(t)\right) \,|\, t \in \mathbb{T} \right\},
\end{equation}
where $\mathbb{T} = \{ t_1, t_2, \ldots, t_N \}$ with $t_k = (k-1)T_s$ for $k \in \mathbb{N}_N$. Here, $T_s$ denotes the sampling time and $f_s = \frac{1}{T_s}$ is the sampling frequency.
\end{defn}


\begin{rem} \label{rem:experimentperf}
The small output $\tilde y(t)$ can be obtained either by smoothing the response to the single input $u_L(t)+\tilde u(t)$ \cite{LTVfNL}, or by running two experiments (with $u(t)=u_L(t)$ and $u(t)=u_L(t)+\tilde u(t)$) and subtracting their outputs \cite{Sadeghifac,JointPaper,Allen3}. Moreover, if $f\in\mathcal C^3$, one may use the symmetric inputs $u_L(t)\pm\tilde u(t)$ and compute 
$
\tilde y(t)=\frac{y_+(t)-y_-(t)}{2}
$, 
which lowers the approximation error in Theorem~\ref{thm:linearapprox} to $\mathcal{O}(\varepsilon^3)$ \cite{ebrahimkhani2025combined}.
\end{rem}


\begin{defn} \label{defn:DFT}
The Discrete Fourier Transform ($\mathrm{DFT}$) of a sampled signal in vectorized form $\mathbf{r} \in \mathbb{R}^N$ is:
\begin{equation}
  \mathrm{DFT}\{\mathbf{r}\} \coloneqq  \mathbf{r}_{\mathcal{F}}= \mathbf{F} \mathbf{r}
\end{equation}
with $\mathbf{F} \in \mathbb{C}^{N \times N}$ denoting the $\mathrm{DFT}$ matrix operator, whose elements are defined as:
\begin{subequations} 
\begin{align}
& F(\xi,t_k) \coloneqq  N^{-\frac{1}{2}} e^{-\frac{j 2 \pi \xi t_k}{N T_s}}, \quad \mathrm{for } \; \xi \in \mathbb{K}, \, t_k \in \mathbb{T} \\
&  \mathbb{K} = \left\{ -\left\lfloor \frac{N}{2} \right\rfloor, -\left\lfloor \frac{N}{2} \right\rfloor + 1, \ldots, \left\lceil \frac{N}{2} \right\rceil - 1 \right\} 
\end{align}
\end{subequations}
where $F(\xi, t_k)$ is the $(\xi, k)$-th element of the matrix $\mathbf{F}$, where $t_k$ corresponds to the $k$-th sample in $\mathbb{T}$.
\end{defn}

\begin{defn} \label{defn:iDFT}
Because the DFT operator $\mathbf{F}$ is unitary (i.e., $\mathbf{F}^H \mathbf{F} = \mathbf{I}_N$), its inverse—the inverse DFT (iDFT)—is defined as:
\begin{equation}
   \mathrm{iDFT}\{ \mathbf{r}_{\mathcal{F}} \} \coloneqq \mathbf{r}= \mathbf{F}^H  \mathbf{r}_{\mathcal{F}} 
\end{equation}
\end{defn}

\begin{assum} \label{assum:BLsignal} 
The spectra of the large and small signals are zero beyond the Nyquist frequency $f_{\mathrm{nyq}}=\frac{f_s}{2}$, i.e. the signals are assumed to be band-limited.
\end{assum}

\begin{rem}
For periodic signals, the time derivative can be expressed by multiplying the DFT by (a power of) $ j\omega_{\xi} $. Here, $ \omega_{\xi} = \frac{2 \pi \xi}{N T_s} $ denotes the angular frequency corresponding to the $ \xi $-th harmonic. This formulation simplifies the handling of time derivatives in the system equations. For non-periodic signals, however, a transient term must be considered \cite{LPVBFE,lataire2011frequency}. In this case, the transient response is a smooth function of $ j\omega_{\xi} $ and can be captured by a polynomial in $ j\omega_{\xi} $ of degree $ \text{max}(n_a+1, n_b) - 1 $ \cite{lataire2011frequency}.
\end{rem}

\begin{assum} \label{assum:noisesmallsig}
The measurements of the input-output small signals are disturbed by additive, zero-mean, stationary noise, i.e.:
\begin{subequations} \label{eq:measnoise}
\begin{align}
\tilde{u}(t) &= \tilde{u}^{\circ}(t) + \nu_u(t) \label{eq:inputmeasnoise} \\
\tilde{y}(t) &= \tilde{y}^{\circ}(t) + \nu_y(t) \label{eq:outputmeasnoise}
\end{align}
\end{subequations}
Here, $\tilde{u}^\circ(t)$ and $\tilde{y}^\circ(t)$ are the noise-free small‑signal input and output, while $\nu_u(t)$ and $\nu_y(t)$ denote their respective measurement noises. These noises are uncorrelated with the true signals $\tilde{u}^\circ(t)$ and $\tilde{y}^\circ(t)$.
\end{assum}

\begin{assum} \label{assum:noiselargesig}
The measurements of the large trajectory of the system $\mathbf{p}_L(t)$ in Definition~\ref{defn:TimeVarOpPoint} are noise-free.
\end{assum}
\begin{rem}
The large trajectory $\mathbf{p}_L$ represents a slowly varying operating point. We assume the Signal-to-Noise Ratio (SNR) for this operating point (or for $u_L(t)$ and $y_L(t)$) is high, rendering measurement noise negligible. Small signals, in contrast, are broadband variations around $\mathbf{p}_L$ with low SNR due to their small amplitudes. Moreover, measurement noise predominantly occurs at higher frequencies \cite[Chapter 12]{NoiseIsHighFq2}.
\end{rem}
Under Assumption~\ref{assum:noisesmallsig}, the $\mathrm{DFT}$s of the input and output small signals can be written as follows:
\begin{subequations}
\begin{align}
& \mathbf{\tilde{u}}_{\mathcal{F}}=\mathbf{\tilde{u}}^{\circ}_{\mathcal{F}}+\bm{\nu}^u_{\mathcal{F}} \\
& \mathbf{\tilde{y}}_{\mathcal{F}}=\mathbf{\tilde{y}}^{\circ}_{\mathcal{F}}+\bm{\nu}^y_{\mathcal{F}}
\end{align}
\end{subequations}
Let  $\mathbf{\tilde{u}}_{\mathcal{F}}=\mathbf{F} \mathbf{\tilde{u}}$, $\tilde{\mathbf{y}}_{\mathcal{F}}=\mathbf{F} \mathbf{\tilde{y}}$,
where $\tilde{\mathbf{u}},\tilde{\mathbf{y}}\in\mathbb{R}^N$ stack $\tilde{u}(t)$  and $\tilde{y}(t)$ for $t\in\mathbb{T}$.  Likewise, $\bm{\nu}^u_{\mathcal{F}}$ and $\bm{\nu}^y_{\mathcal{F}}$ are the DFTs of the noise vectors $\nu_u(t)$ and $\nu_y(t)$, and $\tilde{\mathbf{u}}^\circ_{\mathcal{F}}$, $\tilde{\mathbf{y}}^\circ_{\mathcal{F}}$ denote the noise‑free spectra.
\begin{cor} \label{cor:noiseCov}
The noise covariance matrices are defined as:
\begin{subequations}  \label{eq:noiseCovdef}
\begin{align}
& \mathbf{C}_{\mathcal{F}}^u \coloneqq \mathbb{E} \left\{ \bm{\nu}^u_{\mathcal{F}} (\bm{\nu}^u_{\mathcal{F}})^H  \right\} \\
& \mathbf{C}_{\mathcal{F}}^y \coloneqq \mathbb{E} \left\{ \bm{\nu}^y_{\mathcal{F}} (\bm{\nu}^y_{\mathcal{F}})^H   \right\} \\
& \mathbf{C}_{\mathcal{F}}^{uy} \coloneqq \mathbb{E} \left\{ \bm{\nu}^u_{\mathcal{F}} (\bm{\nu}^y_{\mathcal{F}})^H  \right\}
\end{align}
\end{subequations}
Since the noise is stationary, the covariances in \eqref{eq:noiseCovdef} are asymptotically diagonal matrices. The stationary noise can be represented as filtered white noise, whose DFT is asymptotically independent across frequencies \cite[Theorem 16.25]{SysIDFreqB}.
\end{cor}
\begin{rem}
In some applications, the (co-)variances of $\bm{\nu}^y_{\mathcal{F}}$ and $\bm{\nu}^u_{\mathcal{F}}$ are known a priori. Otherwise, these variances must be estimated from the data. When multiple periods of the input and scheduling signals are available, one can derive a nonparametric noise model in the frequency domain \cite{PINTELON20112892}. For arbitrary inputs with smoothly varying parameters, see \cite{lataire2009estimating}.
\end{rem}
\subsection{LPV System in Frequency Domain} \label{sec:LPVsysFreqDom}
The DFT of the sampled, windowed ($t \in \mathbb{T}$) LPV system \eqref{eq:LPV1} is given by:
\begin{equation} \label{eq:Algebraiceq1}
\bm{\psi}_{n_a+1} \odot \mathbf{\tilde{y}}_{\mathcal{F}}=\mathbf{F} \mathbf{\Omega}_{\mathrm{LPV}} \mathbf{c}_{\mathrm{LPV}}+\mathbf{\Psi} \bm{\gamma} 
\end{equation}
where $\bm{\psi}_n \in \mathbb{C}^{N}$ is the vectorized form of $(j \omega_{\xi})^n$, for $ \xi \in \mathbb{K}$, and:
\begin{subequations} \label{eq:LPVmatest}
\begin{align}
& \mathbf{c}_{\mathrm{LPV}} \coloneqq [\mathbf{c}^\top_1 \ \mathbf{c}^\top_2 \ \ldots \ \mathbf{c}_{n_x}^{\top}]^{\top} \in \mathbb{R}^{N n_x \times 1} \label{eq:LPVcoeffAgebEq}  \\
& \mathbf{\Omega}_{\mathrm{LPV}} \coloneqq  [\mathbf{\Omega}_1 \ \mathbf{\Omega}_2 \ \ldots \ \mathbf{\Omega}_{n_x}] \in \mathbb{R}^{N \times N n_x } \\
& \mathbf{\Omega}_n \coloneqq \begin{cases}
\mathbf{\Omega}_{n-1,\tilde{y}}  & \text{for } 1 \leq n \leq n_a+1 \\
\mathbf{\Omega}_{n-n_a-2,\tilde{u}}  & \text{for } n_a+2 \leq n \leq n_x
\end{cases}
\end{align}
\end{subequations}
where $\mathbf{c}_j \in \mathbb{R}^{N}, j \in \mathbb{N}_{n_x}$ is the vectorized form of the unknown LPV coefficients $c_j(\mathbf{p}_L(t))$ for $t \in \mathbb{T}$, and the diagonal matrix:
\begin{equation}
\mathbf{\Omega}_{n,r} \coloneqq \mathrm{diag} (\mathbf{F}^{H}(\bm{\psi}_n \odot
\mathbf{r}_{\mathcal{F}}) )
\end{equation}
with $r=\tilde{u} \; \mathrm{or} \; \tilde{y} $, and replacing $\mathbf{r}_{\mathcal{F}}$ with the corresponding DFT $\mathbf{\tilde{u}}_{\mathcal{F}} \; \mathrm{or} \; \mathbf{\tilde{y}}_{\mathcal{F}} \in \mathbb{C}^{N}$.
$\mathbf{\Psi} \bm{\gamma}$ represents a polynomial expressed as a vector in terms of $j \omega_{\xi}$ with a degree of $N_{\gamma}$:
\begin{equation}
\mathbf{\Psi} =[\bm{\psi}_0 \  \bm{\psi}_1 \ \ldots \ \bm{\psi}_{N_{\gamma}}] \in \mathbb{C}^{N \times (N_{\gamma}+1) } 
\end{equation}
and $\bm{\gamma} \in \mathbb{R}^{N_{\gamma}+1}$. This polynomial $\mathbf{\Psi} \bm{\gamma}$ takes into account the transient term and alias error  \cite{LTVRKHS,LPVBFE}.
Finally, from the LPV model in the frequency domain in \eqref{eq:Algebraiceq1}, we define the LPV equation error as follows:
\begin{equation}\label{eq:LPVeqErr}
\mathbf{e} \coloneqq -\bm{\psi}_{n_a+1} \odot \mathbf{\tilde{y}}_{\mathcal{F}}+\mathbf{F} \mathbf{\Omega}_{\mathrm{LPV}} \mathbf{c}_{\mathrm{LPV}}+\mathbf{\Psi} \bm{\gamma} 
\end{equation}
The vector $\mathbf{c}_{\mathrm{LPV}}$ is the reshaped version of the vectorized form of $\mathbf{c}(\mathbf{p}_L)$ in \eqref{eq:PV2}, evaluated at the $N$ sample points.

\begin{rem}
The transformation of the continuous‐time ODE \eqref{eq:LPV1} into the algebraic equation \eqref{eq:Algebraiceq1} introduces a residual aliasing error \cite{lataire2011frequency}, particularly for nonperiodic yet band‐limited signals. However, this error is smooth and can be captured by the transient term $\mathbf{\Psi}\bm\gamma$. Consequently, the polynomial degree can be increased by a few orders until the estimation residuals no longer decrease, i.e.\ $N_{\gamma} \geq \mathrm{max}(n_a+1, n_b) - 1$ \cite{LTVRKHS}.
\end{rem}
Note that the frequency‑domain LTV model of \eqref{eq:LPV1} is similar to the LPV model in \eqref{eq:Algebraiceq1}, with the coefficient functions $c_j(\mathbf{p}_L)$ in \eqref{eq:LPVcoeffAgebEq} replaced by the vectorized LTV coefficients $c_j(t)$ for $t\in\mathbb{T}$.

\subsection{Covariance of  Equation Error}
The equation error \eqref{eq:LPVeqErr} can be rewritten as follows:
\begin{equation} \label{eq:Algebraiceq2}
\mathbf{e}=\mathcal{A} \mathbf{\tilde{y}}_{\mathcal{F}} + \mathcal{B}\mathbf{\tilde{u}}_{\mathcal{F}}+\mathbf{\Psi} \bm{\gamma} 
\end{equation}
with:
\begin{subequations}
\begin{align*}
\mathcal{A}
&= -\mathrm{diag}\{\bm\psi_{n_a+1}\}
   + \sum_{n=0}^{n_a}
     \mathbf{F}\,\mathrm{diag}(\mathbf{c}_{n+1})\,\mathbf{F}^H\,
     \mathrm{diag}(\bm\psi_n)\\
\mathcal{B}
&= \sum_{n=0}^{n_b}
    \mathbf{F}\,\mathrm{diag}(\mathbf{c}_{n+n_a+2})\,\mathbf{F}^H\,
    \mathrm{diag}(\bm\psi_n).
\end{align*}
\end{subequations}
Then we have:
\begin{equation} \label{eq:covEwght}
\begin{aligned}
& \mathrm{cov} \{ \mathbf{e} \} \coloneqq 
 \mathbb{E} \left\{  (\mathbf{e} - \mathbb{E} \left\{ \mathbf{e} \right\} )  (\mathbf{e} - \mathbb{E} \left\{ \mathbf{e}  \right\} )^H   \right\} \\
&= \begin{bmatrix}
  \mathcal{A} & \mathcal{B}
\end{bmatrix} \begin{bmatrix}
  \mathbf{C}_{\mathcal{F}}^y &  \mathbf{C}_{\mathcal{F}}^{yu} \\
  \mathbf{C}_{\mathcal{F}}^{uy} &  \mathbf{C}_{\mathcal{F}}^u
\end{bmatrix}
 \begin{bmatrix}
  \mathcal{A}^H \\
  \mathcal{B}^H
\end{bmatrix}
\end{aligned}
\end{equation}

\section{LPV Parametrization}\label{sec:LPVparam}
When considering the LPV model in \eqref{eq:LPV1}, note that its coefficients in \eqref{eq:PV1} are the partial derivatives of the NL  function \eqref{eq:NLsys} evaluated along the nominal trajectory. In other words, the vector field $\mathbf{c}(\mathbf{p}_L)$ is the gradient $\nabla f(\mathbf{p}_L)$. However, not every choice of $\mathbf{c}(\mathbf{p}_L)$ yields a valid gradient field. We therefore present the necessary and sufficient conditions for $\mathbf{c}(\mathbf{p}_L)$ to be a true gradient. 
When the parameterization of the LPV coefficients satisfies  $\mathbf{c} = \nabla f$, the linearization in \eqref{eq:LPV1} holds by construction. Thus, for the dataset \eqref{eq:DataSet0}, the LPV model is always the linearization of the underlying NL system.
\begin{defn}[Curl and Curl-Free Vector Field] \label{def:curlOP}
Let $\mathbf{c} : \mathcal{U} \to \mathbb{R}^{n_x}$ be a vector field such that $\mathbf{c} \in \mathcal{C}^1$. The curl of $\mathbf{c}(\mathbf{p}_L)$ is defined as 
$
\mathrm{Curl}(\mathbf{c}(\mathbf{p}_L)) \coloneqq \nabla \times \mathbf{c}(\mathbf{p}_L)
$, 
where $\times$ denotes the cross product. The vector field $\mathbf{c}(\mathbf{p}_L)$ is called curl-free if $\mathrm{Curl}(\mathbf{c}(\mathbf{p}_L))=0$, which implies:
\begin{equation} \label{eq:CurlfreePDF}
 \frac{\partial c_i ( \mathbf{p}_L)}{\partial p_j}=\frac{\partial c_j ( \mathbf{p}_L)}{\partial p_i},\; \;  (i,j \in \mathbb{N}_{n_x}, \; i \neq j)
\end{equation}
where \eqref{eq:PL} is rewritten as:
\begin{equation}  \label{eq:PLaspi}
 \mathbf{p}_L \coloneqq \begin{bmatrix}
p_{1} & p_{2} & \dots & p_{n_x}
\end{bmatrix} ^\top
\end{equation}
This condition is equivalent to the symmetry of the Jacobian matrix $\nabla \mathbf{c} (\mathbf{p}_L)$ (see \cite[Section 4.1]{KhalilB}).
\end{defn}

\begin{thm}  [Gradient (Section 6 in \cite{VectorCalculus})]\label{thm:GD}
Let $\mathbf{c}: \mathcal{U} \rightarrow \mathbb{R}^{n_x}$, and $\mathbf{c} \in \mathcal{C}^1$. Then, $\forall \mathbf{p}_L \in \mathcal{U}$, $\mathbf{c}(\mathbf{p}_L) = \nabla f(\mathbf{p}_L)$ for some scalar-valued function $f: \mathcal{U} \rightarrow \mathbb{R}$, with $f \in \mathcal{C}^2$ and $\mathcal{U} \subseteq \mathbb{R}^{n_x}$, if and only if $\mathrm{Curl}(\mathbf{c}(\mathbf{p}_L)) = 0$ and $\mathcal{U}$ is a simply connected domain.
\end{thm}



With the conditions for LPV coefficients to form a gradient field established, we now embed them in a vector‑valued RKHS.
\section{Vector-Valued RKHS} \label{sec:vectorVRKHS}
This section introduces the fundamental theoretical aspects of the vector-valued RKHS. The vector of LPV coefficients, $\mathbf{c}(\mathbf{p}_L)$, is modeled within this framework.

\begin{defn}  [Operator-valued kernel \cite{OPkernelDef3,micchelli2005learning}]\label{def:OpVaKernel}
Let $\mathcal{X}$ be a nonempty metric space. A real operator‑valued kernel is a continuous, symmetric, positive semi‑definite map 
 $\mathbf{K}:\mathcal{X}\times\mathcal{X}\to\mathbb{R}^{n_x\times n_x}$.
 Specifically,  $\forall n_N\in\mathbb{N}$, any distinct points $\mathbf{x}_k \in \mathcal{X}$ and $\mathbf{z}_k \in \mathbb{R}^{n_x}$, $k \in \mathbb{N}_{n_N}$: 
$
\sum_{k=1}^{n_N}\sum_{l=1}^{n_N}
\langle \mathbf{z}_k,\,
\mathbf{K}(\mathbf{x}_k,\mathbf{x}_l)\,\mathbf{z}_l
\rangle
\ge0
$. 
\end{defn}
\begin{thm}[Vector-valued RKHS \cite{OPkernelDef3,micchelli2005learning}]\label{def:SVRKHS}
Given an operator‑valued kernel \(\mathbf{K}\) as in Definition~\ref{def:OpVaKernel}, there is a unique smooth vector‑valued RKHS \(\mathcal{H}\) of functions \(\mathbf{h}:\mathcal{X}\to\mathbb{R}^{n_x}\) with reproducing kernel \(\mathbf{K}\) and inner product \(\langle\cdot,\cdot\rangle_{\mathcal{H}}\). Moreover,  $\forall(\mathbf{x},\mathbf{z}) \in \mathcal{X} \times \mathbb{R}^{n_x}$ and  $\forall \mathbf{h} \in \mathcal{H}$, the reproducing property holds:
\begin{equation} \label{eq:ReproProp}
\langle \mathbf{h}(\mathbf{x}),\mathbf{z} \rangle = \langle \mathbf{h}(\cdot),\mathbf{K}_x \mathbf{z} \rangle_{\mathcal{H}}
\end{equation}
where $\mathbf{K}_x \mathbf{z} \coloneqq \mathbf{K}(\cdot,\mathbf{x}) \mathbf{z} \in \mathcal{H}$.
\end{thm}
\begin{defn}  [Curl-free kernel \cite{OPkernelDef3,fuselier2007refined}]\label{def:curlfkernel}
A curl‑free kernel is an operator‑valued kernel \(\mathbf{K}_{\mathrm{curl}} : \mathcal{X} \times \mathcal{X} \to \mathbb{R}^{n_x \times n_x}\) defined as the Hessian of a scalar‑valued stationary kernel:
\begin{equation} \label{eq:HessOfScaK}
\mathbf{K}_{\mathrm{curl}}(\mathbf{x},\mathbf{x}^\prime) \coloneqq    - \nabla_{\mathbf{x}} \nabla_{\mathbf{x}^\prime}^{\top} k_{\sigma}(\mathbf{x},\mathbf{x}^\prime)
\end{equation}
where $k_{\sigma} : \mathcal{X} \times \mathcal{X} \rightarrow \mathbb{R}$ , $\mathcal{X}  \subseteq \mathbb{R}^{n_x}$, is a stationary kernel, and $\nabla_{\mathbf{x}}$ and $\nabla_{\mathbf{x}^\prime}$ denote the gradient operators with respect to $\mathbf{x}$ and $\mathbf{x}^\prime$, respectively. Note that the function $k_{\sigma} (\mathbf{x} - \mathbf{x}^\prime)$ is at least in class  $ \mathcal{C}^2$.
\end{defn}

\begin{thm}[Curl-free RKHS \cite{OPkernelDef3}] \label{thm:curlFRKHS}
Let $\mathcal{H}_{\mathrm{curl}}$ be a vector-valued RKHS with a stationary kernel $\mathbf{K} : \mathcal{X} \times \mathcal{X} \rightarrow \mathbb{R}^{n_x \times n_x}$ as its reproducing kernel. Then, $\mathcal{H}_{\mathrm{curl}}$ is the RKHS of curl-free vector-valued functions, i.e.,  $\mathrm{curl}(\mathbf{h}(\cdot)) = 0$ for all $\mathbf{h} \in \mathcal{H}_{\mathrm{curl}}$ if and only if $\mathrm{curl}(\mathbf{K}_x \mathbf{z}) = 0$ for all $\mathbf{z} \in \mathbb{R}^{n_x}$.
\end{thm}



\begin{rem} \label{rem:kerneis2sSmoother}
If $\mathcal{H}$ is an RKHS with a stationary reproducing kernel, then the kernel is twice as smooth as the functions in the space into which $\mathcal{H}$ is embedded \cite{OPkernelDef3}. Hence, when the curl-free kernel is $\mathbf{K}_{\mathrm{curl}}(\mathbf{x} - \mathbf{x}^\prime) \in \mathcal{C}^2$, the functions in the corresponding RKHS belongs to the class of $\mathcal{C}^1$ functions.
\end{rem}
Selecting $\mathbf{K} = \mathbf{K}_{\mathrm{curl}}$ ensures $\mathrm{curl}(\mathbf{K}_x \mathbf{z}) = 0$,  $\forall \mathbf{z} \in \mathbb{R}^{n_x}$ (see \eqref{eq:HessOfScaK}). Under Theorem~\ref{thm:GD} and Assumption~\ref{assum:simplyconnected}, the  RKHS $\mathcal{H}_{\mathrm{curl}}$ of curl-free functions is well-suited for modeling LPV coefficients. The structural relation \eqref{eq:CurlfreePDF} applies to all functions in $\mathcal{H}_{\mathrm{curl}}$, and using dataset \eqref{eq:DataSet0}, the LPV model represents the linearized version of the unknown NL system \eqref{eq:NLsys}. For estimation, we set $\mathcal{X} = \mathcal{U}$ for LPV coefficients and $\mathcal{X} = \mathbb{T} \subset \mathbb{R}_{\ge 0}$ for LTV.

\section{Proposed Estimators} \label{Sec:PropMeth}
In this section, we propose two sparse estimators in a vector-valued RKHS based on sensitivity measures. By combining sensitivity‐driven sparsity regularization, we select the inputs of the unknown NL function $f(\cdot)$ and identify the scheduling dependencies of the LPV coefficients $c_j(\mathbf{p}_L)$. The resulting reduced LPV model is then used to recover the sparse NL model in Assumption~\ref{assum:SparsityofNLf}.
\begin{thm}[LPV estimator] \label{thm:FreqDomainEstimator}
Consider the LPV model \eqref{eq:Algebraiceq1} without sparsity constraints. Let $\mathcal H_{\mathrm{curl}}$ be the RKHS induced by the curl‐free kernel $\mathbf K_{\mathrm{curl}}:\mathcal X\times\mathcal X\to\mathbb R^{n_x\times n_x}$. The estimation of $\mathbf c\in\mathcal H_{\mathrm{curl}}$ is obtained as:
\begin{equation} \label{eq:argminLPVbasic}
\hat{\mathbf{c}}(\cdot) = \underset{\mathbf{c} \in \mathcal{H}_{\mathrm{curl}}}{\mathrm{argmin}} \, \mathcal{J}_{\mathrm{LPV}}(\mathbf{c})
\end{equation}
with:
\begin{equation} \label{eq:basicCostcfunct}
\mathcal{J}_{\mathrm{LPV}}(\mathbf{c}) = \mathbf{e}^{H} \mathbf{W}^{-1} \mathbf{e} + \gamma_{\mathrm{reg}} \| \mathbf{c} \|^2_{\mathcal{H}_{\mathrm{curl}}}
\end{equation}
where $\mathbf e$ is in \eqref{eq:Algebraiceq2} and $\mathbf W$ is a symmetric, positive definite weighting matrix (see Remark~\ref{rem:Winitialgusconx}).  By the Representer Theorem \cite{micchelli2005learning,scholkopf2001generalized}, the solution admits the finite expansion
\begin{equation} \label{eq:basicReprester}
\hat{\mathbf c}(\cdot)
=\sum_{k=1}^N\mathbf K_{\mathrm{curl}}\bigl(\cdot,\mathbf p_L(t_k)\bigr)\,\bm\alpha_k,
\quad \bm\alpha_k\in\mathbb R^{n_x}.
\end{equation}
\end{thm}
Note that in Theorem~\ref{thm:FreqDomainEstimator}, replacing the curl‐free kernel $\mathbf K_{\mathrm{curl}}$ with the diagonal separable kernel $k_{\sigma}\mathbf I_{n_x}$ \cite{wittwar2018interpolation} treats each LPV coefficient as an independent scalar function (see e.g.\ \cite{LTVRKHS}). In contrast, since the coefficients $c_j$ are the partial derivatives of the NL map $f(\cdot)$ (Definition~\ref{defn:Linearapprox}), the curl‐free kernel $\mathbf K_{\mathrm{curl}}$ and its RKHS $\mathcal H_{\mathrm{curl}}$ more naturally encode this gradient structure.


If the vector field $\mathbf{c}$ admits a representer as in \eqref{eq:basicReprester}, then the $j$-th  LPV coefficient can be expressed as follows:
\begin{equation} \label{eq:PVrepresent}
\hat{c}_j(\cdot)=\sum_{k=1}^{N}  \mathbf{K}_{\mathrm{curl}}(\cdot,\mathbf{p}_L(t_k))_{j,:} \bm{\alpha}_k
\end{equation}
with $\mathbf{K}_{\mathrm{curl}}(\cdot,\mathbf{p}_L(t_k))_{j,:}$ as the $j$-th row of $\mathbf{K}_{\mathrm{curl}}(\cdot,\mathbf{p}_L(t_k))$. 
\begin{defn}[Sensitivity functions] \label{defn:Sensivityvecmax}
The sensitivity vector (or matrix) of the scalar function $f(\cdot)$  in \eqref{eq:NLsys} (or vector-valued function $\mathbf{c}$ in \eqref{eq:PV2}) evaluated at the trajectory $\mathbf{p}_L$ are defined as:
\begin{subequations}
\begin{align}
\mathbf{s}(\mathbf{p}_L) &\coloneqq \nabla f (\mathbf{p}_L) \in \mathbb{R}^{n_x} \label{eq:Sensivityvector} \\
\mathbf{S}(\mathbf{p}_L) &\coloneqq \nabla \mathbf{c}(\mathbf{p}_L) \in \mathbb{R}^{n_x \times n_x} \label{eq:sensivitymatrix}
\end{align}
\end{subequations}
\end{defn}
Note that per Definition~\ref{defn:Sensivityvecmax}, the LTV/LPV coefficients reflect the sensitivity of $f(\cdot)$. The Jacobian matrix of $\mathbf{c}(\mathbf{p}_L)$, denoted $\mathbf{S}(\mathbf{p}_L)$, acts as the sensitivity matrix for these coefficients and matches the Hessian of $f(\cdot)$ at $\mathbf{p}_L$. When $\mathbf{c}(\mathbf{p}_L)$ is curl-free, $\mathbf{S}(\mathbf{p}_L)$ is symmetric.



The significance of each input variable $p_l, l \in \mathbb{N}_{n_x}$ (in \eqref{eq:PLaspi}), on the function $h(\cdot)$ (where $h(\cdot)$ denotes either $f(\cdot)$ or $c_j(\cdot)$) is quantified by the magnitude of its partial derivative $\frac{\partial h(\cdot)}{\partial p_l}$. If $p_l$ does not influence $h(\cdot)$, this derivative is zero \cite{bai2019variable,mukherjee2006learning,rosasco2013nonparametric}. Since $\frac{\partial h(\cdot)}{\partial p_l}$ measures the rate of change of $h(\cdot)$ with respect to $p_l$, it motivates selecting input variables based on a norm of these derivatives. In the following subsections, we apply this approach to select input variables for both the NL function $f(\cdot)$ and the LPV coefficients $c_j(\cdot)$.

\begin{rem}
Sensitivity is a local measure and may vanish even if an input remains relevant—e.g., when the domain  $\mathcal{U}$ is disconnected. If  $h$ is piecewise constant on disjoint regions of  $\mathcal{U}$, then  $\partial h/\partial p_l=0$  everywhere, although  $p_l$ is still pertinent (see \cite{rosasco2013nonparametric}). In this work, we assume that $\partial h/\partial p_l=0$ (for $h=f$ or $c_j$) implies that  $h$ is constant with respect to $p_l$.
\end{rem}


\subsection{NL System Input Variable Selection}  \label{Sec:NLinputSelect}
In this section, we address input‐variable selection for the NL map in \eqref{eq:NLsys}. Noting that the LPV/LTV coefficients in \eqref{eq:LPV1} form the sensitivity vector in \eqref{eq:Sensivityvector}, input selection for $f$ reduces to model‐order selection of the LPV approximation. We thus propose a sparse RKHS estimator for LPV model‐order selection.
We use regularization to enforce sparsity in the LPV coefficients vector \(\mathbf{c}(\cdot)\), which minimizes the number of non-zero elements (corresponding to the \(\ell_0\)-norm). Since computing the \(\ell_0\)-norm is NP-hard in higher dimensions \cite{weston2003use}, we employ the \(\ell_1\)-norm, its tightest convex relaxation \cite{rosasco2013nonparametric}. Following \cite{laurain2020sparse}, we apply the \(\ell_1\)-norm to the \(\ell_{\infty}\)-norms of the LPV coefficients, adding this regularization term to \eqref{eq:basicCostcfunct} for LPV model order selection:
\begin{equation} \label{eq:linfregultermLPVord}
\mathcal{S}_{\mathrm{ord}}(\mathbf{c}) = \left\| 
\begin{array}{c}
\left[ \|c_1\|_{\infty}, \|c_2\|_{\infty}, \ldots, \|c_{n_x}\|_{\infty} \right] 
\end{array}
\right\|_{1}
\end{equation} 
Therefore, we formulate the following optimization problem to estimate the LPV coefficients and enforce sparsity in the model order:
\begin{equation} \label{eq:CostFuncLPVmdlOrder}
\begin{aligned}
 \hat{\mathbf{c}}(\cdot)=
 \underset{\substack{\mathbf{c} \in \mathcal{H}_{\text{curl}} }}{\text{argmin}} \mathcal{J}_{\mathrm{curl}}(\mathbf{c}) + \gamma_{\mathrm{ord}} \mathcal{S}_{\mathrm{ord}}(\mathbf{c})
\end{aligned}
\end{equation}
Here, $\mathcal{J}_{\mathrm{LPV}}(\mathbf{c})$ is defined in \eqref{eq:basicCostcfunct}, and $\gamma_{\mathrm{ord}}>0$ weights the sparsity penalty. The equivalent epigraph form \cite{boyd2004convex} of the optimization problem in \eqref{eq:CostFuncLPVmdlOrder} is:
\begin{equation} \label{eq:argminLPVmdlOrder}
\begin{aligned}
 \hat{\mathbf{c}}(\cdot)=
 \underset{\substack{\mathbf{c} \in \mathcal{H}_{\text{curl}}, \; \rho_j >0  \\  -\rho_j \leq c_j(\mathbf{p}_L) \leq \rho_j, \mathbf{p}_L \in \mathcal{U}}}{\text{argmin}} \mathcal{J}_{\mathrm{LPV}}(\mathbf{c}) + \gamma_{\mathrm{ord}} \sum_{j=1}^{n_x} \rho_j
\end{aligned}
\end{equation}
with $\rho_j$ as slack variables. The problem \eqref{eq:argminLPVmdlOrder} is an optimization problem with an infinite number of constraints.
In the following, we employ a sampling-based (pointwise) relaxation of the functional constraints in \eqref{eq:argminLPVmdlOrder}. This approach is motivated by the fact that the space $\mathcal{H}_{\text{LPV}}$ consists of smooth, vector-valued functions. Intuitively, enforcing the constraints in \eqref{eq:argminLPVmdlOrder} at a finite set of grid points is sufficient. To convert the functional constraints into sampling-based constraints, we define the $n_t$ nodal (time) points as:
\begin{equation} \label{eq:nodalPoints}
\mathbb{T}' \coloneqq \{ t'_1, t'_2, \ldots, t'_{n_t} \}
\end{equation} 
 Then, the  $\ell_{\infty}$-norm is approximated as follows \cite{laurain2020sparse}:
\begin{equation} \label{eq:linfapproxorder}
\|c_j\|_{\infty} \approx \max\limits_{t' \in \mathbb{T}'} \left| \vphantom{\frac{c_j}{\mathbf{p}_L(t')}} c_j(\mathbf{p}_L(t')) \right|
\end{equation}
Hence, \eqref{eq:argminLPVmdlOrder} becomes:
\begin{equation} \label{eq:argminLPVmdlOrderFinite_dim_relax}
\begin{aligned}
 \hat{\mathbf{c}}(\cdot)=
 \underset{\substack{\mathbf{c} \in \mathcal{H}_{\text{curl}}, \; \rho_j > 0 \\  -\rho_j \leq c_j(\mathbf{p}_L(t')) \leq \rho_j, t' \in \mathbb{T}'}}{\text{argmin}} \mathcal{J}_{\mathrm{LPV}}(\mathbf{c}) + \gamma_{\mathrm{ord}} \sum_{j=1}^{n_x} \rho_j
\end{aligned}
\end{equation} 
\begin{rem}
The points $\mathbb{T}'$ can be chosen by time gridding, random selection (e.g., \cite{laurain2020sparse,singh2021learning}), directly from the sampled data (i.e., $\mathbb{T}' = \mathbb{T}$) (e.g., \cite{gregorova2018structured,10990161}), or as a strict superset of the measurement points  $\mathbb{T}$ \cite{singh2021learning} to cover a wider operational range; however, this last approach requires measuring the large trajectory for the additional points.
\end{rem}

\begin{thm}[Sparsity in LPV model order] \label{thm:LPVmodelOrderRepresenter}
Consider a curl-free kernel $\mathbf{K}_{\mathrm{curl}}$  in Definition~\ref{def:curlfkernel} and the associated curl-free RKHS $\mathcal{H}_{\mathrm{curl}}$ in Definition~\ref{thm:curlFRKHS}. We define the finite-dimensional subspace:
\begin{equation*} \label{eq:RepreseLPVModelOrder}
\mathcal{V}_{\mathrm{ord}}  \coloneqq \left\{  \sum_{i=1}^{N_{\mathrm{ord}}}  \mathbf{K}_{\mathrm{curl}}(\cdot,\mathbf{p}_L(\check{t}_i)) \bm{\alpha}_i, \; \bm{\alpha}_i \in \mathbb{R}^{n_x}  \right\} \subset \mathcal{H}_{\mathrm{curl}}
\end{equation*}
where  $\check{t}_i$ is the $i$-th element of  $\mathbb{T}_{\mathrm{ord}} = \mathbb{T} \cup \mathbb{T}'$, and  $N_{\mathrm{ord}}$ is the cardinality of $\mathbb{T}_{\mathrm{ord}}$.
 Then, if $\hat{\mathbf{c}}(\cdot)$ is the minimizer of \eqref{eq:argminLPVmdlOrderFinite_dim_relax}, then $\hat{\mathbf{c}}(\cdot) \in \mathcal{V}_{\mathrm{ord}}$.
\end{thm}
\begin{pf}
Since $\mathcal{V}_{\mathrm{ord}}$ is a finite-dimensional subspace of $\mathcal{H}_{\mathrm{curl}}$, it is closed. Hence, by the Projection Theorem \cite[Chapter 3]{dullerud2013course}: $\mathcal{H}_{\mathrm{curl}} = \mathcal{V}_{\mathrm{ord}} \oplus \mathcal{V}^{\perp}_{\mathrm{ord}}$, where $\mathcal{V}^{\perp}_{\mathrm{ord}}$ is the orthogonal complement of $\mathcal{V}_{\mathrm{ord}}$. Then, any $\mathbf{c} \in \mathcal{H}_{\mathrm{curl}}$ can be written in a unique way as $\mathbf{c} = \mathbf{c}^{\parallel} + \mathbf{c}^{\perp}$ with $\mathbf{c}^{\parallel} \in \mathcal{V}_{\mathrm{ord}}$, $\mathbf{c}^{\perp} \in \mathcal{V}^{\perp}_{\mathrm{ord}}$, and:
\begin{equation} \label{eq:Innerprodperparallz}
\langle \mathbf{c}^{\parallel}, \mathbf{c}^{\perp} \rangle_{\mathcal{H}_{\mathrm{curl}}} = 0
\end{equation}
The cost function in \eqref{eq:argminLPVmdlOrderFinite_dim_relax} consists of three terms:

\textbf{(I)} The data fit term (the first term in \eqref{eq:basicCostcfunct}) depends only on the function $\mathbf{c}$ through its evaluation at the measurement points $\mathbb{T}$. By using the reproducing property \eqref{eq:ReproProp}, $\forall \mathbf{z} \in \mathbb{R}^{n_x}$, $\forall t \in \mathbb{T}_{\mathrm{ord}}$:
\begin{equation} \label{eq:costandimprictermIND}
\begin{aligned}
& \langle \mathbf{c}(\mathbf{p}_L(t)), \mathbf{z} \rangle =  \langle \mathbf{c}(\cdot), \mathbf{K}_{\mathrm{curl}}(\cdot, \mathbf{p}_L(t)) \mathbf{z} \rangle_{\mathcal{H}_{\mathrm{curl}}} = \\  
& \langle \mathbf{c}^{\parallel}(\cdot) + \mathbf{c}^{\perp}(\cdot), \mathbf{K}_{\mathrm{curl}}(\cdot, \mathbf{p}_L(t)) \mathbf{z} \rangle_{\mathcal{H}_{\mathrm{curl}}} = \\  
& \langle \mathbf{c}^{\parallel}(\cdot), \mathbf{K}_{\mathrm{curl}}(\cdot, \mathbf{p}_L(t)) \mathbf{z} \rangle_{\mathcal{H}_{\mathrm{curl}}}
\end{aligned}
\end{equation}
 since $\mathbf{K}_{\mathrm{curl}}(\cdot, \mathbf{p}_L(t)) \mathbf{z} \in \mathcal{V}_{\mathrm{ord}}$. Thus, this term in the cost function is independent of $\mathbf{c}^{\perp}(\cdot)$. 

\textbf{(II)} For the constraint in \eqref{eq:argminLPVmdlOrderFinite_dim_relax}, this term depends only on the function $\mathbf{c}$ through its evaluation at the nodal points $\mathbb{T}'$. By using the same method as in \eqref{eq:costandimprictermIND} through the reproducing property, $\mathbf{c}^{\perp}(\cdot)$ plays no role in the pointwise relaxation of the constraints in \eqref{eq:argminLPVmdlOrderFinite_dim_relax}, and hence does not affect the minimizer. 

\textbf{(III)} For the last (regularization) term in \eqref{eq:argminLPVmdlOrderFinite_dim_relax}, we have:
\begin{equation*}
\| \mathbf{c} \|^2_{\mathcal{H}_{\mathrm{curl}}} = \| \mathbf{c}^{\parallel} + \mathbf{c}^{\perp} \|^2_{\mathcal{H}_{\mathrm{curl}}} = \| \mathbf{c}^{\parallel} \|^2_{\mathcal{H}_{\mathrm{curl}}} + \| \mathbf{c}^{\perp} \|^2_{\mathcal{H}_{\mathrm{curl}}}
\end{equation*}
which is obtained by applying \eqref{eq:Innerprodperparallz}. Therefore, $\mathbf{c}^{\perp} = 0$, and the optimal $\mathbf{c}$ lies in $\mathcal{V}_{\mathrm{ord}}$.
\end{pf}
 


\begin{cor}[Optimization problem] \label{cor:optprobmdlord}
Model order selection for the LPV model \eqref{eq:LPV1} can be formulated as a quadratic program. The matrix‐form optimization problem appears in Appendix~\ref{appx:LPVmdlordrOptprob}.
\end{cor}
\begin{rem}
Note that, for model‐order selection, one can use the LTV representation in place of the LPV model. In this case, Theorem~\ref{thm:LPVmodelOrderRepresenter} still applies once we replace the curl‐free kernel by the diagonal separable kernel  $k_{\sigma}(\mathcal{X},\mathcal{X}')\,\mathbf I_{n_x},
   \mathcal{X} = \mathbb{T}\subset\mathbb{R}_{\ge0}$. However, the structural relationship in \eqref{eq:CurlfreePDF} is then lost.
\end{rem}
\begin{rem} \label{rem:Winitialgusconx}
If the weighting matrix $\mathbf W$ is constant, Corollary~\ref{cor:optprobmdlord} reduces to a convex quadratic program. However, in \cite{LTVRKHS}, the weighting matrix $\mathbf{W}$ is taken to be the diagonal matrix with diagonal entries  $\text{diag}(\mathrm{cov} \{ \mathbf{e} \})$ (from \eqref{eq:covEwght}). Although the kernel‐based estimator is not consistent, this choice reduces the mean‐squared error. Moreover, when the system belongs to the model class \eqref{eq:LPV1}, the minimiser of the expected data‐fit term $\mathbf e^H\mathbf W^{-1}\mathbf e$ in \eqref{eq:argminLPVmdlOrderFinite_dim_relax} is independent of the noise properties \cite{LTVRKHS}. In that case $\mathbf W$ depends on the unknown parameters, yielding a nonconvex quadratic program. We therefore employ the convex relaxation of \cite{LTVRKHS} (Algorithm~\ref{alg:LPVSchedulingRed}), using as an initial guess the solution obtained with $\mathbf W=\mathbf I_N$.
\end{rem}
If $p_j$ does not contribute to $f$, then $\frac{\partial f}{\partial p_j}=c_j=0$.
Hence, whenever $\hat{c}_j=0$, $p_j$ can be removed from the LPV scheduling vector by deleting its associated row and column in $\mathbf K_{\mathrm{curl}}$.

\subsection{Reduction of Scheduling Dependency} \label{Sec:LPVschSelect}
In this section, we estimate the LPV coefficients  $c_j(\mathbf p_L)$ and their scheduling dependencies via the sensitivity measure from Section~\ref{Sec:NLinputSelect}. We estimate the gradient $\mathbf c(\mathbf p_L)$ in \eqref{eq:Sensivityvector} while imposing sparsity on the sensitivity matrix \eqref{eq:sensivitymatrix}. Since $f\in\mathcal C^2$, each $c_j\in\mathcal C^1$. We then employ the curl‐free kernel $\mathbf K_{\mathrm{curl}}(\mathbf x-\mathbf x')\in\mathcal C^2$, whose RKHS $\mathcal H_{\mathrm{curl}}$ consists of $\mathcal C^1$ functions (Remark~\ref{rem:kerneis2sSmoother}).
 The entries of \eqref{eq:sensivitymatrix} are the partial derivatives of the LPV coefficients $c_j(\mathbf{p}_L)$ with respect to each scheduling variable $p_l$. To enforce sparsity, we introduce the regularization term:
\begin{equation} \label{eq:l1regulschselection}
\mathcal{S}_{\mathrm{sch}}(\mathbf{c}) = \sum_{j=1}^{n_x} \left\| \left\|\frac{\partial c_j}{\partial p_1}\right\|_{\infty} \left\|\frac{\partial c_j}{\partial p_2}\right\|_{\infty} \ldots \left\|\frac{\partial c_j}{\partial p_{n_x}}\right\|_{\infty} \right\|_{1}
\end{equation}
This involves applying the $\ell_1$-norm to the $\ell_\infty$-norms of the entries in the sensitivity matrix  \eqref{eq:sensivitymatrix}. Although the number of regularization terms in  \eqref{eq:l1regulschselection} can be large, parameterizing the vector field $\mathbf{c}$ as a gradient field ensures its Jacobian is a symmetric Hessian matrix. This symmetry \eqref{eq:CurlfreePDF} reduces the number of regularization terms. Additionally, removing any input to $f(\cdot)$ (Section~\ref{Sec:NLinputSelect}) eliminates the corresponding row and column from $\mathbf{S}$.
Then, $\mathcal{S}_{\mathrm{sch}}(\mathbf{c}) $ can be simplified as follows:
\begin{equation} \label{eq:l1regulschselectionreduced}
\mathcal{S}_{\mathrm{sch}}(\mathbf{c}) = \sum_{j=1}^{n_x} \sum_{l=j}^{n_x} \left\|\frac{\partial c_j}{\partial p_l}\right\|_{\infty}
\end{equation}
This reduction removes $\frac{n_x(n_x-1)}{2}$ regularization terms from the original $n_x n_x$ in \eqref{eq:l1regulschselection}. In \cite{rosasco2013nonparametric}, the $\ell_1$-norm over the $\ell_2$-norm of the partial derivatives is employed to address the regularization problem. However, in this section, we adopt the approach described in Section~\ref{Sec:NLinputSelect}, using \eqref{eq:l1regulschselectionreduced} to promote sparsity in the input variables of the LPV coefficients.
We then formulate the optimization problem to jointly estimate the LPV coefficients and select the scheduling variables as follows:
\begin{equation} \label{eq:CostFuncLPVschselect1}
\begin{aligned}
 \hat{\mathbf{c}}(\cdot) = \underset{\substack{\mathbf{c} \in \mathcal{H}_{\text{curl}}}}{\text{argmin}} \; \mathcal{J}_{\mathrm{LPV}}(\mathbf{c}) + \gamma_{\mathrm{sch}}\, \mathcal{S}_{\mathrm{sch}}(\mathbf{c})
\end{aligned}
\end{equation}
where $\gamma_{\mathrm{sch}}>0$ is the regularization weight. The equivalent epigraph form \cite{boyd2004convex} of \eqref{eq:CostFuncLPVschselect1} is given by:
\begin{equation} \label{eq:CostFuncLPVschselect_epigraphForm}
\begin{aligned}
 \hat{\mathbf{c}}(\cdot) = \underset{\substack{\mathbf{c} \in \mathcal{H}_{\text{curl}}, \; \tau_{j,l} > 0 \\ -\tau_{j,l} \leq \frac{\partial c_j(\mathbf{p}_L)}{\partial p_l} \leq \tau_{j,l}, \; \mathbf{p}_L \in \mathcal{U}}}{\text{argmin}} \; \mathcal{J}_{\mathrm{LPV}}(\mathbf{c}) + \gamma_{\mathrm{sch}} \sum_{j=1}^{n_x} \sum_{l=j}^{n_x} \tau_{j,l},
\end{aligned}
\end{equation}
with $\tau_{j,l}$ being additional optimization variables. Note that \eqref{eq:CostFuncLPVschselect_epigraphForm} represents an optimization problem with an infinite number of constraints. To render the problem tractable, we adopt the same pointwise relaxation as in Section~\ref{Sec:NLinputSelect} using the nodal (time) points defined in \eqref{eq:nodalPoints}; hence, we approximate:
\begin{equation} \label{eq:linfapproxsch}
\left\| \frac{\partial c_j}{\partial p_l} \right\|_{\infty} \approx \max_{t' \in \mathbb{T}'} \left| \frac{\partial c_j (\mathbf{p}_L(t'))}{\partial p_l} \right|.
\end{equation}
Finally, the estimation of the vector field $\mathbf{c}$ and the selection of LPV scheduling dependencies is obtained by solving the following problem:
\begin{equation} \label{eq:CostFuncLPVschselect_epigraphForm_relaxation}
\begin{aligned}
 \hat{\mathbf{c}}(\cdot)=
 \underset{\substack{\mathbf{c} \in \mathcal{H}_{\text{curl}}, \; \tau_{j,l} > 0   \\  -\tau_{j,l} \leq\frac{\partial c_j(\mathbf{p}_L(t'))}{\partial p_l} \leq \tau_{j,l}, t' \in \mathbb{T}' }}{\text{argmin}} \mathcal{J}_{\mathrm{LPV}}(\mathbf{c}) + \gamma_{\mathrm{sch}} \sum_{j=1}^{n_x} \sum_{l=j}^{n_x} \tau_{j,l}
\end{aligned}
\end{equation}
\begin{thm}[Sparsity in LPV coefficients input] \label{thm:LPVschvariableRepresenter}
Consider a curl-free kernel $\mathbf{K}_{curl}$ in Definition~\ref{def:curlfkernel} with $\mathbf{K}_{\mathrm{curl}}(\mathbf{x} - \mathbf{x}^\prime) \in \mathcal{C}^2$, and the associated curl-free RKHS $\mathcal{H}_{curl}$. Then, we define the finite-dimensional subspace:
\begin{equation*} \label{eq:RepreseLPVSchselec}
\begin{aligned}
& \mathcal{V}_{\mathrm{sch}}  \coloneqq \left\{  \sum_{k=1}^{N}  \mathbf{K}_{\mathrm{curl}}(\cdot,\mathbf{p}_L(t_k)) \bm{\alpha}_k + \right.\\
& \left.  \sum_{l=1}^{n_x}  \sum_{s=1}^{n_t}  \frac{\partial   \mathbf{K}_{\mathrm{curl}}(\cdot,\mathbf{p}_L(t'_s))}{\partial p_l}  \bm{\alpha}'_{l,s}, \; \bm{\alpha}_k,\bm{\alpha}'_{l,s} \in \mathbb{R}^{n_x}  \right\} \subset \mathcal{H}_{\mathrm{curl}}
\end{aligned}
\end{equation*}
where $t_k \in \mathbb{T}$ and $t'_s \in \mathbb{T}'$. Denote $\hat{\mathbf{c}}(\cdot)$ as the optimizer for the problem~\eqref{eq:CostFuncLPVschselect_epigraphForm_relaxation}, then $\hat{\mathbf{c}}(\cdot) \in  \mathcal{V}_{\mathrm{sch}}$. ($\frac{\partial   \mathbf{K}_{\mathrm{curl}}(\cdot,\mathbf{p}_L(t'_s))}{\partial p_l}$ is the elementwise partial derivative of $\mathbf{K}_{\mathrm{curl}}(\cdot,\mathbf{p}_L)$ with respect to $p_l$ and evaluated at $t'_s$).
\end{thm}
\begin{pf}
Similar to the proof of Theorem~\ref{thm:LPVmodelOrderRepresenter}, since $\mathcal{V}_{\mathrm{sch}}$ is a finite-dimensional subspace of $\mathcal{H}_{\mathrm{curl}}$, it is closed. Hence, by the Projection Theorem \cite[Chapter 3]{dullerud2013course}: $\mathcal{H}_{\mathrm{curl}}=\mathcal{V}_{\mathrm{sch}} \oplus \mathcal{V}^{\perp}_{\mathrm{sch}}$, where $\mathcal{V}^{\perp}_{\mathrm{sch}}$ is the orthogonal complement of $\mathcal{V}_{\mathrm{sch}}$. Then any $\mathbf{c} \in \mathcal{H}_{\mathrm{curl}}$ can be written as $\mathbf{c}=\mathbf{c}^{ \parallel}+\mathbf{c}^{ \perp}$ with $\mathbf{c}^{\parallel} \in \mathcal{V}_{\mathrm{sch}}$ and $\mathbf{c}^{\perp} \in \mathcal{V}^{\perp}_{\mathrm{sch}}$.
 The cost function  in \eqref{eq:CostFuncLPVschselect_epigraphForm_relaxation} consists of three terms: 

\textbf{(I)} for the data fit term (the first term in \eqref{eq:basicCostcfunct}) (similar to the first part of the proof of Theorem~\ref{thm:LPVmodelOrderRepresenter}), this term only depends on the function $\mathbf{c}$ through the evaluation at the measurement points  $\mathbb{T}$. Then, by using the reproducing property, it is independent of $\mathbf{c}^{ \perp}$ (see \eqref{eq:costandimprictermIND}).

\textbf{(II)} For the partial derivative constraints in \eqref{eq:CostFuncLPVschselect_epigraphForm_relaxation}, this term depends on $\mathbf{c}$ through the evaluation of its partial derivatives evaluated at the grid points $\mathbb{T}'$. Then, by using the derivative properties of $\mathcal{H}_{\mathrm{curl}}$ \cite{OPkernelDef3} for the partial derivative of $c_j$ with respect to $p_l$, we have:
\begin{equation}
\begin{aligned}
&\frac{\partial c_j(\mathbf{p}_L(t'_s))}{\partial p_l}  = \langle  \frac{\partial \mathbf{c}(\mathbf{p}_L(t'_s))}{\partial p_l},\mathbf{e}_j \rangle =\\
&  \langle \mathbf{c}(\cdot),\frac{\partial   \mathbf{K}_{\mathrm{curl}}(\cdot,\mathbf{p}_L(t'_s))}{\partial p_l}  \mathbf{e}_j \rangle_{\mathcal{H}_{\mathrm{curl}}} = \\
& \langle \mathbf{c}^{ \parallel}(\cdot)+\mathbf{c}^{ \perp}(\cdot),\frac{\partial   \mathbf{K}_{\mathrm{curl}}(\cdot,\mathbf{p}_L(t'_s))}{\partial p_l} \mathbf{e}_j \rangle_{\mathcal{H}_{\mathrm{curl}}}=\\
& \langle \mathbf{c}^{ \parallel}(\cdot),\frac{\partial   \mathbf{K}_{\mathrm{curl}}(\cdot,\mathbf{p}_L(t'_s))}{\partial p_l} \mathbf{e}_j \rangle_{\mathcal{H}_{\mathrm{curl}}}
\end{aligned}
\end{equation}
because $\frac{\partial   \mathbf{K}_{\mathrm{curl}}(\cdot,\mathbf{p}_L(t'_s))}{\partial p_l} \mathbf{e}_j \in \mathcal{V}_{\mathrm{sch}}$, and $\mathbf{e}_j$ being the $j$-th basis coordinate in $\mathbb{R}^{n_x}$. This means that the constraints are independent of $\mathbf{c}^{\perp}$; hence, $\mathbf{c}^{\perp}$ plays no role in the pointwise relaxation of the constraints in \eqref{eq:CostFuncLPVschselect_epigraphForm_relaxation} and  does not affect the minimizer.

\textbf{(III)} For the last (regularization) term we have:
\begin{equation}
\| \mathbf{c} \|^2_{\mathcal{H}_{\mathrm{curl}}}=\| \mathbf{c}^{ \parallel}+\mathbf{c}^{ \perp} \|^2_{\mathcal{H}_{\mathrm{curl}}}=\| \mathbf{c}^{ \parallel} \|^2_{\mathcal{H}_{\mathrm{curl}}} + \| \mathbf{c}^{ \perp} \|^2_{\mathcal{H}_{\mathrm{curl}}}
\end{equation}
This is obtained by applying \eqref{eq:Innerprodperparallz}. Therefore, $\mathbf{c}^{ \perp} =0$, and the optimal $\mathbf{c}$ lies in $\mathcal{V}_{\mathrm{sch}}$.

\end{pf}

\begin{rem}
The representers used for scheduling dependency selection and LPV model order reduction are curl‐free, since 
$\mathcal{V}_{\mathrm{sch}},\mathcal{V}_{\mathrm{ord}}\subset\mathcal{H}_{\mathrm{curl}}$. 
Hence, any reduction in the dependency of one coefficient $c_j\in\mathbf{c}$ on $p_l$ enforces a matching reduction in another coefficient via the curl‐free constraint \eqref{eq:CurlfreePDF}. 
\end{rem}
\begin{cor}[Optimization problem] \label{corr:LPVschselec}
Selecting the input variables for the LPV coefficients $c_j, j \in \mathbb{N}_{n_x}$ can be cast as a quadratic program. The matrix‐form optimization problem is given in Appendix~\ref{appx:LPVschreducOptprob}.
\end{cor}


\begin{rem}
Input and scheduling variable selection can be formulated by combining the cost functions  
\eqref{eq:argminLPVmdlOrderFinite_dim_relax} and  
\eqref{eq:CostFuncLPVschselect_epigraphForm_relaxation}.  
 However, the two estimators are presented independently and can be executed sequentially using Algorithm~\ref{alg:LPVSchedulingRed} for model order and scheduling selection.
\end{rem}
\begin{rem}
The kernel $\mathbf{K}_{\mathrm{curl}}$ may be any stationary, twice continuously differentiable curl‑free kernel. Its selection depends on the system’s physics and operating range: for example, known local periodicity or polynomial nonlinearities suggests a corresponding $k_{\sigma}$ in \eqref{eq:HessOfScaK}, while smooth NL variations are often well captured by the Squared Exponential (SE) kernel. Finally, the matrix‑valued curl‑free kernel is obtained via \eqref{eq:HessOfScaK}.
\end{rem}         
\begin{rem}
When input variables are highly correlated, multicollinearity can compromise model interpretability. To address this, we refer to \cite{gregorova2018structured}, which proposes a grouped‐lasso‐like regularizer and an elastic‐net penalty for input variable selection via a sparse regularization problem in the RKHS framework. These regularizations can be applied to the proposed $\ell_{\infty}$‐norm in Sections~\ref{Sec:NLinputSelect} and~\ref{Sec:LPVschSelect} in the presence of highly correlated inputs.
\end{rem}

\begin{alg}{LPV model order  (\textbf{or} scheduling)  selection}
\label{alg:LPVSchedulingRed}
\vspace{0.5em} 
\hrule 
\begin{algorithmic}[1] 
\State \textbf{Input:} LPV model structure \eqref{eq:LPV1},  grid points $\mathbb{T}'$, curl-free kernel in Definition~\ref{thm:curlFRKHS}, noise covariance in Corollary~\ref{cor:noiseCov}, a threshold $\kappa_o$ for output approximation error.

 \Repeat:
	\State Select $\tilde{u}(t) \in S_\varepsilon$.
\State Run the experiment from Remark~\ref{rem:experimentperf} to obtain $\tilde{y}(t)$ and the dataset (Definition~\ref{defn:DataSet}).
\State Solve the optimization problem in Corollary~\ref{cor:optprobmdlord} for model order (\textbf{or}~\ref{corr:LPVschselec} for scheduling selection), with $\mathbf{W}=\mathbf{I}_N$ for $\hat{\mathbf{c}}$.
    \State \textbf{do}
\State \quad Compute the data-fit term: $\mathcal{J}_{\mathrm{WLS}}^{-} = \mathbf{e}^{H} \mathbf{W}^{-1} \mathbf{e}$.
\State \quad Compute the weighting matrix: $\mathbf{W} = \text{diag}(\mathrm{cov} \{ \mathbf{e} \})$.
\State \quad Solve the optimization problem in Corollary~\ref{cor:optprobmdlord} (\textbf{or}~\ref{corr:LPVschselec}) for $\hat{\mathbf{c}}$.
\State \quad Compute new data-fit term: $\mathcal{J}_{\mathrm{WLS}} = \mathbf{e}^{H} \mathbf{W}^{-1} \mathbf{e}$.
\State \textbf{while} $\mathcal{J}_{\mathrm{WLS}}^{-} \leq \mathcal{J}_{\mathrm{WLS}}$
\State Compute the error in output approximation:
 \Statex $\kappa_y \coloneqq   \sup_{t \geq 0} \left \lVert \hat{\tilde{y}}(t)-\tilde{y}(t) \right \rVert$.
\If{$\kappa_y >  \kappa_o$}
    \State The error is too large, reduce $\varepsilon$.
\EndIf
\Until{$\delta_y < \delta_o$}
    \State \textbf{return} Estimated $\hat{\mathbf{c}}$, and the model orders based on Corollary \ref{cor:optprobmdlord} (\textbf{or} scheduling dependency based on Corollary \ref{corr:LPVschselec}). 
\end{algorithmic}
\end{alg}
\section{NL Model Structure Detection} \label{Sec:NLmdlreduced}
Consider the NL function in \eqref{eq:NLsys}. Under Assumption~\ref{assum:SparsityofNLf}, which decomposes $ f(\cdot) $ into $ M $ independent subfunctions $ \{f_i(\mathbf{p}_i)\}_{i=1}^M $ with disjoint input spaces $ \{\mathcal{U}_i\}_{i=1}^M $, the Hessian matrix of $ f(\cdot) $ exhibits a block diagonal structure. This follows because the second-order partial derivatives between variables from distinct blocks $ \mathbf{p}_i $ and $ \mathbf{p}_j $ ($ i \neq j $) vanish due to the functional independence of $ f_i(\cdot) $ and $ f_j(\cdot) $. Specifically, since the Hessian of $ f(\cdot) $ coincides with the sensitivity matrix $ \mathbf{S} $ defined in \eqref{eq:sensivitymatrix}, Assumption~\ref{assum:SparsityofNLf} ensures that $ \mathbf{S} $ can be expressed as:
\begin{equation} \label{eq:HessianofSparsef}
\mathbf{S}
= \operatorname{diag}\bigl(\mathbf{S}_{1}, \dots, \mathbf{S}_{M}\bigr)
\end{equation}
where each submatrix $ \mathbf{S}_i = \nabla_{\mathbf{p}_i}^2 f_i(\mathbf{p}_i) $ corresponds to the Hessian of the $ i $-th subfunction $ f_i(\cdot) $ with respect to its dedicated variables $ \mathbf{p}_i \in \mathcal{U}_i $. The block diagonal structure reflects the input space separability of the subfunctions: variables in $ \mathbf{p}_i $ do not influence $ f_j(\cdot) $ for $ j \neq i $, and vice versa. Consequently, the second-order cross-derivatives $ \partial^2 f / \partial p_i \partial p_j $ ($p_i \in \mathbf{p}_i, p_j \in \mathbf{p}_j$ and $ i \neq j $) are identically zero.
 After the LPV reduction in Sections~\ref{Sec:NLinputSelect} and~\ref{Sec:LPVschSelect}, the reduced sensitivity matrix in \eqref{eq:sensivitymatrix} keeps the block-diagonal structure of \eqref{eq:HessianofSparsef}. Each block $\mathbf{S}_i$ measures the sensitivity of the reduced LPV model coefficients to changes in the parameter subset $\mathbf{p}_i$. This model represents the sparse NL system at the operating point $\mathbf{p}_L$. The reduced curl-free kernel in Section~\ref{Sec:LPVschSelect} maintains the same block-diagonal sparsity as \eqref{eq:HessianofSparsef}, due to the Hessian's symmetry preserved under the sparsity constraints of Assumption~\ref{assum:SparsityofNLf}.

\section{Simulation Results} \label{sec:Simres}
To illustrate the proposed method, consider the NL system with the sparse additive structure of Assumption~\ref{assum:SparsityofNLf}:
\begin{equation} \label{eq:EXmainmodel}
\begin{aligned}
& \ddot{y}(t) = -y(t) + \frac{1}{2}\Bigl(1 - y^2(t)\Bigr)\dot{y}(t) + 2u(t) + \sin\bigl(2u(t)\bigr) \\
& + \frac{1}{5}\ddot{u}(t) = f_1\bigl(y(t),\dot{y}(t)\bigr) + f_2\bigl(u(t)\bigr) + f_3\bigl(\ddot{u}(t)\bigr)
\end{aligned}
\end{equation}
Starting with an overparameterized NL model for \eqref{eq:EXmainmodel}:
\begin{equation} \label{eq:simres0NLmodel}
\ddot{y}(t) = f\Bigl(y(t), \dot{y}(t), u(t), \dot{u}(t), \ddot{u}(t), \dddot{u}(t)\Bigr),
\end{equation}
our goal is to recover the model structure in Assumption~\ref{assum:SparsityofNLf} by identifying the number of subfunctions $M=3$ and their input spaces for each $f_i(\mathbf{p}_i)$, where $i \in \mathbb{N}_M$. 
\subsection{Simulation setup}
\begin{figure}
\begin{center}
\includegraphics[height=6cm]{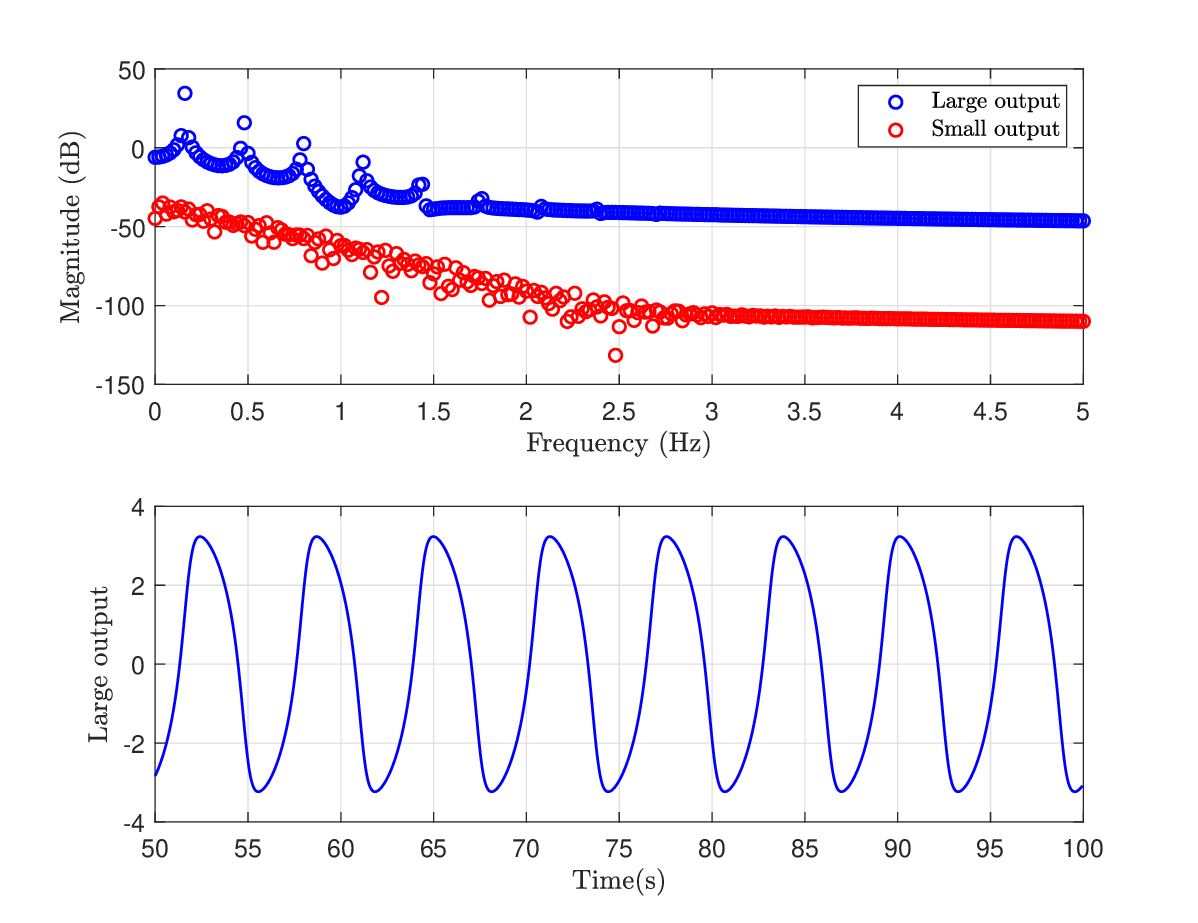} 
\caption{Top: Frequency-domain representation of the large and small output signals. Bottom: Time-domain representation of the large output signal.}
\label{fig1} 
\end{center} 
\end{figure}

For the simulations, we use a large input $u_L(t) = \sin\left(t + \frac{\pi}{4}\right)$,
and a small multisine input with root mean square~$10^{-3}$, fundamental frequency $f_0 = 0.02\,$Hz, and 20 excited harmonics spanning $[f_0,0.4]\,$Hz. The sampling frequency is $f_s = 20\,$Hz.  We employ the multisine signal to precisely target the frequency band of interest.
 The large operating point is perturbed by the small signal. For estimation, we use the interval $[50,100]$s with $N=1000$ samples. Stationary output noise  is modeled by a second‑order discrete filter (resonance $\approx0.174\,$Hz; see Corollary~\ref{cor:noiseCov}) and an SNR of 25 dB.
 In Figure~\ref{fig1} (top), the spectra of the large and small output signals are shown, while the bottom part displays the large output signal in the time domain. The large output signal's spectrum reveals the system's nonlinearity through multiple harmonics (integer multiples of the excitation frequency).
For LPV systems, the output spectrum includes frequencies not directly excited. These extra harmonics, at integer multiples of the scheduling frequency ($\omega_{\mathrm{sch}}$), appear around the excited input frequencies ($\omega_{\mathrm{input}}$), specifically at $\omega_{\mathrm{y}} = \omega_{\mathrm{input}} \pm k\,\omega_{\mathrm{sch}}$ for $k \in \mathbb{N}$ \cite{LPVBFE}.
We evaluate each sparse estimator from Sections~\ref{Sec:NLinputSelect} and~\ref{Sec:LPVschSelect} using Monte Carlo simulations with $N_{\mathrm{mc}} = 50$ independent noise realizations.


\subsection{NL system input variable selection} \label{simres:lpvmdlorder_sel}
\begin{figure}
\begin{center}
\includegraphics[height=6cm]{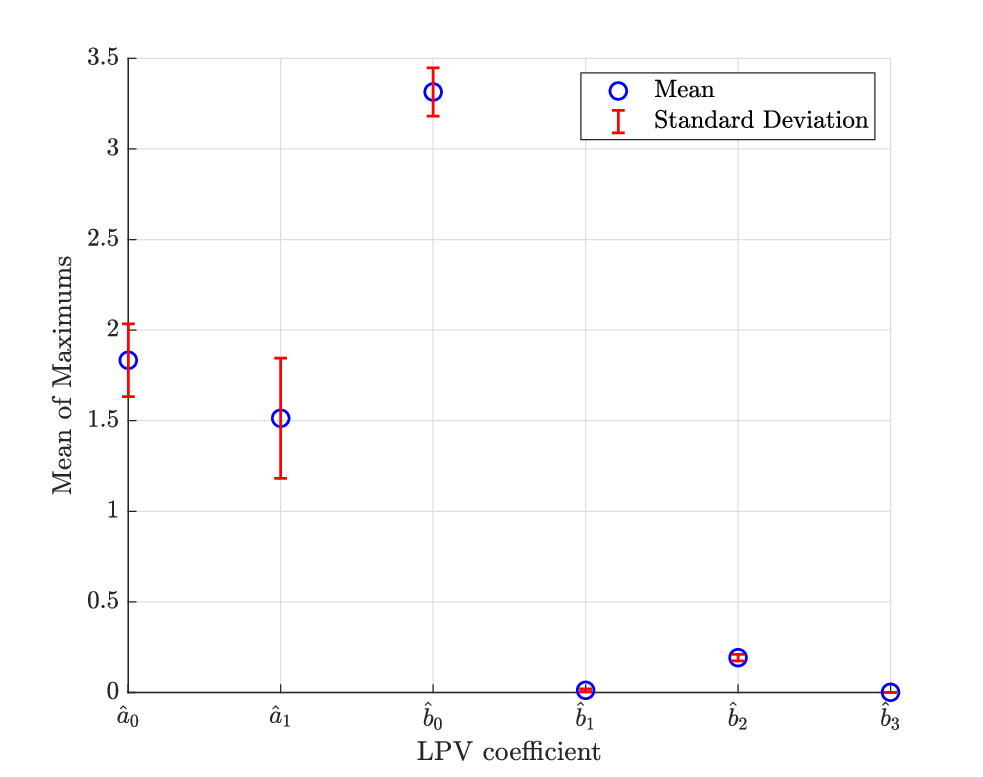} 
\caption{Model order selection for the LPV system  \eqref{simresults_LPV_init}. Blue circles represent the mean of the maximum values of each LPV coefficient over $N_{\mathrm{mc}}  = 50$ Monte Carlo runs, while red bars indicate the corresponding standard deviations. For $\hat{b}_1$, the mean is 0.00119 with a standard deviation of 0.0074 (true value: 0). For $\hat{b}_2$, the mean is 0.1922 with a standard deviation of 0.0177 (true value: 0.2), and for $\hat{b}_3$, the mean is 0.00111 with a standard deviation of 0.0002 (true value: 0).}
\label{fig2} 
\end{center} 
\end{figure}

In this section, we use the method from Section~\ref{Sec:NLinputSelect} to select the order of the LPV model corresponding to the overparameterized NL model \eqref{eq:simres0NLmodel}, given by:
\begin{equation} \label{simresults_LPV_init}
\ddot{\tilde{y}}(t) = \sum_{n=0}^{1} a_{n}(\mathbf{p}_L(t)) \frac{d^{n}\tilde{y}_0(t)}{dt^{n}} + \sum_{m=0}^{3} b_{m}(\mathbf{p}_L(t)) \frac{d^{m}\tilde{u}(t)}{dt^{m}}
\end{equation}
with the scheduling vector 
$
\mathbf{p}_L = \begin{bmatrix}
p_{1} & p_{2} & p_{3} & p_{4} & p_{5} & p_{6}
\end{bmatrix}^\top =
\begin{bmatrix}
y_{L} & \dot{y}_{L} & u_{L} & \dot{u}_{L} & \ddot{u}_{L} & \dddot{u}_{L}
\end{bmatrix}^\top
$. 
The elements of $\mathbf{p}_L$ generally differ in both units and scale. 
When generating the curl‑free kernel in Definition~\ref{def:curlfkernel}, we employ an Automatic Relevance Determination (ARD) SE kernel because it assigns a distinct length‑scale to each input dimension:
$
k_{\sigma}(\mathbf{p}_L,\mathbf{p}'_L)=\exp\left(-\sum_{i=1}^{n_x} \frac{\bigl(p_i - p'_i\bigr)^2}{2\sigma_i^2}\right)
$. 
Here, $p_i$ and $p'_i$ denote the $i$th components of $\mathbf{p}_L$ and $\mathbf{p}'_L$. We choose lengthscales $\bigl[\sigma_1, \dots, \sigma_6\bigr] = [7,2,7,7,7,7]$
as reasonable values within the search space. An infinite lengthscale in any dimension implies the function is constant along that axis \cite{piironen2016projection}.
We set the regularization parameter in \eqref{eq:basicCostcfunct} to $\gamma_{\mathrm{reg}} = 10^{-3}$ for all subsequent simulations.
Various hyperparameter tuning methods are reviewed in \cite{RKHSgnr2}, and cross-validation approaches for LTV/LPV estimations \cite{laurain2020sparse,LTVRKHS}; note that cross-validation requires an additional validation dataset. 
 We select $n_t =50$ nodal  points in \eqref{eq:nodalPoints} from the measured trajectory $\mathbf{p}_L(t)$, evenly spaced at 1-second intervals.
For stable numerical implementation of the LPV  estimator, we replace the monomials $\bm{\psi}_n=(j\omega_\xi)^n$ (Section \ref{sec:LPVsysFreqDom}) with an orthogonal Legendre basis in $j\omega_\xi$:
\begin{equation}
\bm{\psi}_n(j \omega_{\xi} ) =
\begin{cases} 
    j\, P_n\!\left(\frac{\omega_{\xi}}{\omega_{\mathrm{M}}}\right) & \text{if } n \text{ is odd} \\[1mm]
    P_n\!\left(\frac{\omega_{\xi}}{\omega_{\mathrm{M}}}\right) & \text{if } n \text{ is even}
\end{cases}
\end{equation}
where $\omega_M=\max_{\xi\in\mathbb K}\omega_\xi$ and $P_n$ is the $n$th‐degree Legendre polynomial. This scaling prevents large powers of $j\omega_\xi$ and balances magnitudes across a broad frequency range \cite{ebrahimkhani2025combined}.
To balance the regularization scales in \eqref{eq:CostFuncLPVmdlOrder}, we normalize the constraints in \eqref{eq:argminLPVmdlOrderFinite_dim_relax}.  Each LPV coefficient $c_j = \frac{\partial f}{\partial p_j}$, is scaled by  $\frac{p_{j\mathrm{M}}}{f_{\mathrm{M}}}$, where $p_{j\mathrm{M}} = \max_{t \in \mathbb{T}} |p_j(t)|$ and $f_{\mathrm{M}}$ is the maximum of $f(\cdot)$ over the operational range, yielding dimensionless, sensitivity‑weighted terms.  For model‑order selection, since $f(\cdot)$ is common to all sensitivities, we set $f_{\mathrm{M}}=1$. This acts as a tuning parameter for the regularization term $\gamma_{\mathrm{ord}}$. We then solve the optimization problem in Corollary~\ref{cor:optprobmdlord}.
 Figure~\ref{fig2} shows the mean and uncertainty bounds of the maximum values of the estimated LPV coefficients.  The results indicate $\hat b_1=\hat b_3=0$, so $f(\cdot)$ is independent of those inputs.  Since the sparse estimator’s extra regularization introduces bias, we recommend re‑estimating via Theorem~\ref{thm:FreqDomainEstimator} after removing $\hat b_1$ and $\hat b_3$.
\subsection{LPV scheduling selection}
In this section, we sparsify the Hessian matrix (the sensitivity matrix \eqref{eq:sensivitymatrix}) of the unknown NL function $ f(\cdot) $. We use the same kernel as before and discard the LPV coefficients $ \hat{b}_1 $ and $ \hat{b}_3 $, eliminating the scheduling variables $ p_4 $ and $ p_6 $. This results in a reduced scheduling vector $ \mathbf{p}_L = [p_1, p_2, p_3, p_5]^\top $, with lengthscales $ [\sigma_1, \sigma_2, \sigma_3, \sigma_5] = [7, 2, 7, 7] $, and a corresponding reduced curl-free kernel.
 The regularization in  \eqref{eq:l1regulschselection} initially includes 16 terms of the form $ \frac{\partial c_j}{\partial p_l} $ for $ j, l \in \{1, 2, 3, 5\} $. By exploiting the Hessian’s symmetry, we reduce this to 10 terms (see  \eqref{eq:l1regulschselectionreduced}), considering $ \frac{\partial c_j}{\partial p_l} $ only for $ j \leq l $, which retains the main diagonal and lower triangular elements.
 We use the same nodal points \(\mathbb{T}'\) as before. The constraints in  \eqref{eq:CostFuncLPVschselect_epigraphForm} reflect the sensitivity of each LPV coefficient $ c_j(\cdot) $ to scheduling variable $ p_l $.  We normalize these sensitivities as $ \frac{\partial c_j}{\partial p_l} \frac{p_{l\mathrm{M}}}{c_{j\mathrm{M}}} $, where $ p_{l\mathrm{M}} $ is the maximum value of $ p_l $  in the operational range. Here, $ c_{j\mathrm{M}} $ is the maximum of $ c_j $ from the reduced model, derived via Theorem~\ref{thm:FreqDomainEstimator}.

After solving the optimization problem in  Corollary~\ref{corr:LPVschselec} over $N_{\mathrm{mc}} = 50$ Monte Carlo runs, we compute the mean of each element in the sensitivity matrix \eqref{eq:sensivitymatrix} at the grid points $\mathbb{T}'$. Figure~\ref{fig3} shows these means along with their uncertainty bounds. The shaded subfigures (upper triangular off-diagonal elements) are excluded from regularization. They are plotted for illustration and mirror the lower triangular elements.
Figure~\ref{fig3} reveals that: 
$ \hat{a}_0 $ depends on $ p_1 $ and $ p_2 $,
$ \hat{a}_1 $ depends only on $ p_1 $,
$ \hat{b}_0 $ depends only on $ p_3 $ with no interaction with other coefficients,
$ \hat{b}_2 $ is constant.
These dependencies are linked via the curl-free constraint in  \eqref{eq:CurlfreePDF}. The reduced LPV model is:
\begin{equation} \label{simres:reducedLPV}
  \ddot{\tilde{y}}(t) = a_0 (p_1,p_2) \dot{\tilde{y}}(t) + a_1 (p_1) \tilde{y}(t) + b_0(p_3) \tilde{u}(t) + \theta_{p_5} \ddot{\tilde{u}}(t)
\end{equation}
where $\theta_{p_5}$ is a constant parameter to be estimated.  
Note that based on Figure~\ref{fig3}, the coefficient $a_1$ is only a function of $p_1$. However, due to the curl-condition \eqref{eq:CurlfreePDF}, since $a_0(p_1,p_2)$ depends on both $p_1$ and $p_2$, the LPV coefficients $a_0$ and $a_1$ can be estimated as two interrelated functions. We employ the  block diagonal curl-free kernel  $
  \mathbf{K}_{\mathrm{r}} (\mathbf{p}_L,\mathbf{p}'_L) = 
 \operatorname{diag}\bigl(\nabla_{\mathbf{x}} \nabla_{\mathbf{x}^\prime}^{\top} k_{\sigma}([p_1,p_2],[p'_1,p'_2]), k_{\sigma}(p_3,p'_3)   \bigr)
$ to estimate the reduced LPV model \eqref{simres:reducedLPV}. 
Note that $b_2$ is a constant (see Figure~\ref{fig3}) and is not modeled as a function in RKHS. 
Figure~\ref{fig4} plots the mean and uncertainty bounds of the final reduced LPV model.

\subsection{NL model structure}
\begin{figure}
\begin{center}
\includegraphics[height=7cm]{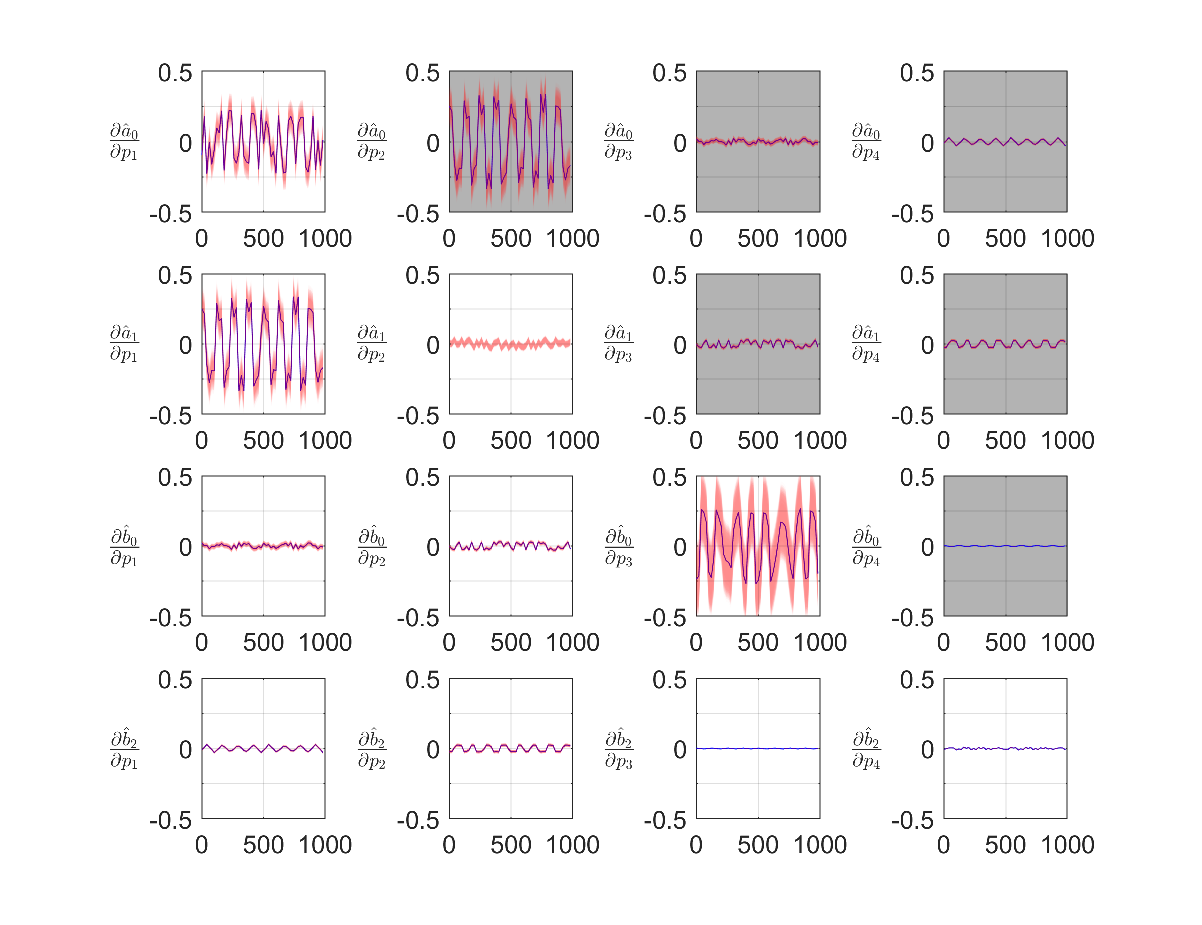} 
\caption{Mean (blue line) and uncertainty bounds (light red region) of each element of the sensitivity matrix \eqref{eq:sensivitymatrix}, estimated over $N_{\mathrm{mc}} = 50 $ Monte Carlo runs and evaluated at the grid points $ \mathbb{T}' $. X‑axis shows sample indices. Shaded (gray) subfigures are omitted due to symmetry in the Hessian matrix \eqref{eq:sensivitymatrix} and are shown for illustration purposes only.}
\label{fig3} 
\end{center} 
\end{figure}

\begin{figure}
\begin{center}
\includegraphics[height=6cm]{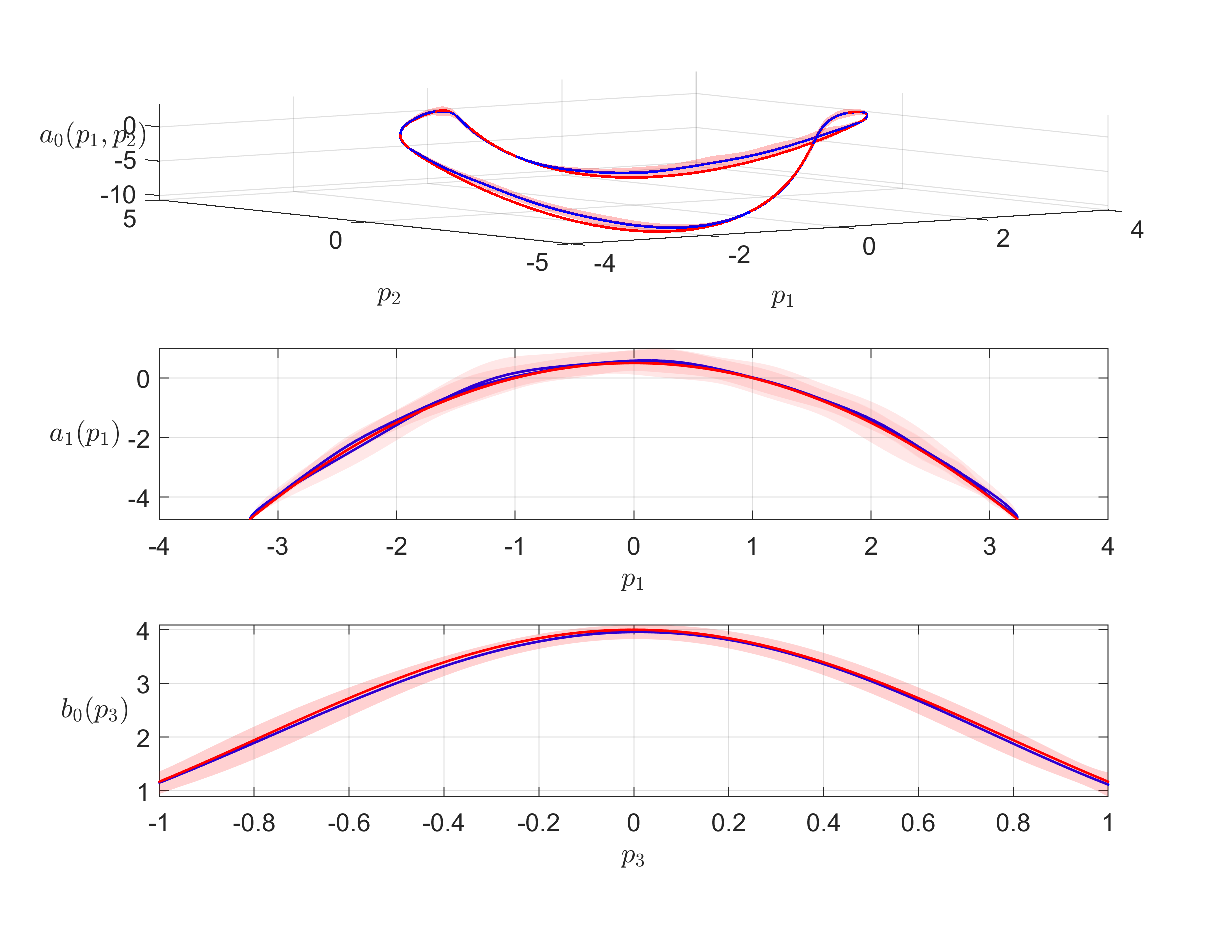} 
\caption{Mean (blue line) and uncertainty bounds (light red shaded region) for each LPV coefficient after model reduction, over  $ N_{\mathrm{mc}} = 50 $ Monte Carlo runs. The true LPV coefficients are superimposed as red lines.}
\label{fig4} 
\end{center} 
\end{figure}

After reducing the LPV model, we identify the NL model structure \eqref{eq:simres0NLmodel} using the method from Section~\ref{Sec:NLmdlreduced}. From Section~\ref{simres:lpvmdlorder_sel}, we determine that $f$ is independent of $p_4$ and $p_6$, so $f = f(p_1, p_2, p_3, p_5)$. Figure~\ref{fig3} shows the Hessian \eqref{eq:sensivitymatrix} of $f(p_1,p_2,p_3,p_5)$ over the operational range $\mathbf{p}_L$, which has a block‑diagonal form:
$ 
\mathbf{S}
= \operatorname{diag}\bigl(\mathbf{S}_{1}, \mathbf{S}_{2}, 0\bigr)
$. 
 Here, $\mathbf{S}_1\in\mathbb{R}^{2\times2}$, $\mathbf{S}_2\in\mathbb{R}^{1\times1}$. This Hessian structure implies that, over the range~$\mathbf{p}_L$, the NL function follows the model structure:
$
  f(p_1, p_2, p_3, p_5) = f_1(p_1, p_2) + f_2(p_3) + \theta_{p_5} p_5
$.
\section{Conclusion} \label{sec:conclu}
We propose a kernel‑based method for reducing NL models via linearization. Perturbation data around varying operating points are approximated by an LPV/LTV model. We introduce a model structure for LPV coefficients, with necessary and sufficient conditions ensuring the LPV model consistently represents the linearized NL system. The sparse structure of the NL system is obtained via LPV model reduction. Within an RKHS framework, two sensitivity‑driven sparse estimators enable selection of LPV model order and coefficient dependencies via regularization. Future work will explore joint RKHS‑based estimation of LPV and NL models through tailored kernel design.


\bibliographystyle{plain}        
\bibliography{REF_file}           

\begin{thebibliography}{10}

\bibitem{ashraf2024partial}
Waqar~Muhammad Ashraf and Vivek Dua.
\newblock Partial derivative-based dynamic sensitivity analysis expression for
  non-linear auto regressive with exogenous (narx) modelcase studies on
  distillation columns and model's interpretation investigation.
\newblock {\em Chemical Engineering Journal Advances}, 18:100605, 2024.

\bibitem{NoiseIsHighFq2}
Karl~Johan {\AA}str{\"o}m and Richard~M Murray.
\newblock {\em Feedback systems: an introduction for scientists and engineers}.
\newblock Princeton university press, 2008.

\bibitem{bai2019variable}
Er-Wei Bai, Changming Cheng, and Wen-Xiao Zhao.
\newblock Variable selection of high-dimensional non-parametric nonlinear
  systems by derivative averaging to avoid the curse of dimensionality.
\newblock {\em Automatica}, 101:138--149, 2019.

\bibitem{billings2013nonlinear}
Stephen~A Billings.
\newblock {\em Nonlinear system identification: NARMAX methods in the time,
  frequency, and spatio-temporal domains}.
\newblock John Wiley \& Sons, 2013.

\bibitem{boyd2004convex}
Stephen Boyd and Lieven Vandenberghe.
\newblock {\em Convex optimization}.
\newblock Cambridge university press, 2004.

\bibitem{CBs2}
Tommaso Bradde, Stefano Grivet-Talocia, Giuseppe~C. Calafiore, Anton~V.
  Proskurnikov, Zohaib Mahmood, and Luca Daniel.
\newblock Bounded input dissipativity of linearized circuit models.
\newblock {\em IEEE Transactions on Circuits and Systems I: Regular Papers},
  67(6):2064--2077, 2020.

\bibitem{CBs1}
Tommaso Bradde, Stefano Grivet-Talocia, Pedro Toledo, Anton~V. Proskurnikov,
  Alessandro Zanco, Giuseppe~C. Calafiore, and Paolo Crovetti.
\newblock Fast simulation of analog circuit blocks under nonstationary
  operating conditions.
\newblock {\em IEEE Transactions on Components, Packaging and Manufacturing
  Technology}, 11(9):1355--1368, 2021.

\bibitem{breiman1995better}
Leo Breiman.
\newblock Better subset regression using the nonnegative garrote.
\newblock {\em Technometrics}, 37(4):373--384, 1995.

\bibitem{brunton2016discovering}
Steven~L Brunton, Joshua~L Proctor, and J~Nathan Kutz.
\newblock Discovering governing equations from data by sparse identification of
  nonlinear dynamical systems.
\newblock {\em Proceedings of the national academy of sciences},
  113(15):3932--3937, 2016.

\bibitem{chang2010electrochemical}
Byoung-Yong Chang and Su-Moon Park.
\newblock Electrochemical impedance spectroscopy.
\newblock {\em Annual Review of Analytical Chemistry}, 3(1):207--229, 2010.

\bibitem{chen2017kernel}
Jianbo Chen, Mitchell Stern, Martin~J Wainwright, and Michael~I Jordan.
\newblock Kernel feature selection via conditional covariance minimization.
\newblock {\em Advances in neural information processing systems}, 30, 2017.

\bibitem{cheng2021variable}
Changming Cheng and Erwei Bai.
\newblock Variable selection based on squared derivative averages.
\newblock {\em Automatica}, 127:109491, 2021.

\bibitem{VectorCalculus}
S.J Colley.
\newblock {\em Vector Calculus}.
\newblock Pearson, Boston, 4 edition, 2011.

\bibitem{corradini2015digital}
Luca Corradini, Dragan Maksimovic, Paolo Mattavelli, and Regan Zane.
\newblock {\em Digital control of high-frequency switched-mode power
  converters}.
\newblock John Wiley \& Sons, 2015.

\bibitem{DSR1}
CA~Desoer and KK~Wong.
\newblock Small-signal behavior of nonlinear lumped networks.
\newblock {\em Proceedings of the IEEE}, 56(1):14--22, 1968.

\bibitem{DSR2}
Charles~A Desoer and Mathukumalli Vidyasagar.
\newblock {\em Feedback systems: input-output properties}.
\newblock SIAM, 2009.

\bibitem{dreesen2015decoupling}
Philippe Dreesen, Mariya Ishteva, and Johan Schoukens.
\newblock Decoupling multivariate polynomials using first-order information and
  tensor decompositions.
\newblock {\em SIAM Journal on Matrix analysis and Applications},
  36(2):864--879, 2015.

\bibitem{dullerud2013course}
Geir~E Dullerud and Fernando Paganini.
\newblock {\em A course in robust control theory: a convex approach},
  volume~36.
\newblock Springer Science \& Business Media, 2013.

\bibitem{Sadeghifac}
Sadegh Ebrahimkhani and John Lataire.
\newblock Identification of nonlinear systems via lpv model identification
  around a time-varying trajectory.
\newblock {\em IFAC-PapersOnLine}, 2023.
\newblock IFAC WORLD CONGRESS 2023.

\bibitem{ebrahimkhani2025combined}
Sadegh Ebrahimkhani and John Lataire.
\newblock A combined estimator for nonlinear system identification via lpv
  approximations.
\newblock {\em International Journal of Robust and Nonlinear Control}, 2025.
\newblock Accepted 7 May 2025.

\bibitem{ebrahimkhani2025continuous}
Sadegh Ebrahimkhani and John Lataire.
\newblock Continuous-time nonlinear system identification and model reduction
  via linearization around a varying operating point.
\newblock {\em Authorea}, May 2025.

\bibitem{10990161}
Sadegh Ebrahimkhani and John Lataire.
\newblock Kernel-based estimation of frequency response function of strictly
  passive systems.
\newblock {\em IEEE Control Systems Letters}, 9:162--167, 2025.

\bibitem{engelbrecht2001new}
Andries~P Engelbrecht.
\newblock A new pruning heuristic based on variance analysis of sensitivity
  information.
\newblock {\em IEEE transactions on Neural Networks}, 12(6):1386--1399, 2001.

\bibitem{fuselier2007refined}
Edward~J Fuselier~Jr.
\newblock {\em Refined error estimates for matrix-valued radial basis
  functions}.
\newblock PhD thesis, Texas A\&M University, 2007.

\bibitem{OPkernelDef3}
Joan Glaunes and Mario Micheli.
\newblock Matrix-valued kernels for shape deformation analysis. geometry.
\newblock {\em Imaging and Computing}, 1(1):57--139, 2014.

\bibitem{LPVBFE}
Jan Goos, John Lataire, Ebrahim Louarroudi, and Rik Pintelon.
\newblock Frequency domain weighted nonlinear least squares estimation of
  parameter-varying differential equations.
\newblock {\em Automatica}, 75:191--199, 2017.

\bibitem{gregorova2018structured}
Magda Gregorova, Alexandros Kalousis, and St{\'e}phane Marchand-Maillet.
\newblock Structured nonlinear variable selection.
\newblock In {\em Proceedings of Conference on Uncertainty in Artificial
  Intelligence (UAI)}. 6-10 August 2018, 2018.

\bibitem{hastie2015statistical}
Trevor Hastie, Robert Tibshirani, and Martin Wainwright.
\newblock Statistical learning with sparsity.
\newblock {\em Monographs on statistics and applied probability}, 143(143):8,
  2015.

\bibitem{hess2014respiratory}
Dean~R Hess.
\newblock Respiratory mechanics in mechanically ventilated patients.
\newblock {\em Respiratory care}, 59(11):1773--1794, 2014.

\bibitem{keymolen2023low}
Andy Keymolen, Antoine Marchal, Frank Heck, Ben Van~Den Elshout, Gerd
  Vandersteen, Joop Jonckheer, and John Lataire.
\newblock Low-frequency respiratory oscillometric measurements during
  mechanical ventilation.
\newblock In {\em 2023 IEEE International Symposium on Medical Measurements and
  Applications (MeMeA)}, pages 1--6, 2023.

\bibitem{KhalilB}
H.~K. Khalil.
\newblock {\em Nonlinear Systems}.
\newblock Prentice Hall, Upper Saddle River, third edition, 2008.

\bibitem{khosravi2023kernel}
Mohammad Khosravi and Roy~S Smith.
\newblock Kernel-based identification with frequency domain side-information.
\newblock {\em Automatica}, 150:110813, 2023.

\bibitem{koelewijn2020scheduling}
Patrick~JW Koelewijn and Roland T{\'o}th.
\newblock Scheduling dimension reduction of lpv models-a deep neural network
  approach.
\newblock In {\em 2020 American Control Conference (ACC)}, pages 1111--1117.
  IEEE, 2020.

\bibitem{koltchinskii2010sparsity}
Vladimir Koltchinskii and Ming Yuan.
\newblock Sparsity in multiple kernel learning.
\newblock {\em The Annals of Statistics}, 38(6):3660--3695, 2010.

\bibitem{kukreja2006least}
Sunil~L Kukreja, Johan L{\"o}fberg, and Martin~J Brenner.
\newblock A least absolute shrinkage and selection operator (lasso) for
  nonlinear system identification.
\newblock {\em IFAC proceedings volumes}, 39(1):814--819, 2006.

\bibitem{kwiatkowski2008pca}
Andreas Kwiatkowski and Herbert Werner.
\newblock Pca-based parameter set mappings for lpv models with fewer parameters
  and less overbounding.
\newblock {\em IEEE Transactions on Control Systems Technology},
  16(4):781--788, 2008.

\bibitem{lai2019sparse}
Zhilu Lai and Satish Nagarajaiah.
\newblock Sparse structural system identification method for nonlinear dynamic
  systems with hysteresis/inelastic behavior.
\newblock {\em Mechanical Systems and Signal Processing}, 117:813--842, 2019.

\bibitem{lataire2009estimating}
John Lataire and Rik Pintelon.
\newblock Estimating a nonparametric colored-noise model for linear slowly
  time-varying systems.
\newblock {\em IEEE Transactions on Instrumentation and Measurement},
  58(5):1535--1545, 2009.

\bibitem{lataire2011frequency}
John Lataire and Rik Pintelon.
\newblock Frequency-domain weighted non-linear least-squares estimation of
  continuous-time, time-varying systems.
\newblock {\em IET Control Theory \& Applications}, 5(7):923--933, 2011.

\bibitem{LTVfNL}
John Lataire, Rik Pintelon, and Tom Oomen.
\newblock An ltv approach to identifying nonlinear systems-with application to
  an rrr-robot.
\newblock {\em IFAC-PapersOnLine}, 54(7):445--450, 2021.
\newblock 19th IFAC Symposium on System Identification SYSID 2021.

\bibitem{LTVRKHS}
John Lataire, Rik Pintelon, Dario Piga, and Roland T{\'o}th.
\newblock Continuous-time linear time-varying system identification with a
  frequency-domain kernel-based estimator.
\newblock {\em IET Control Theory \& Applications}, 11(4):457--465, 2017.

\bibitem{laurain2020sparse}
Vincent Laurain, Roland T{\'o}th, Dario Piga, and Mohamed Abdelmonim~Hassan
  Darwish.
\newblock Sparse {RKHS} estimation via globally convex optimization and its
  application in lpv-io identification.
\newblock {\em Automatica}, 115:108914, 2020.

\bibitem{lee2024small}
Young-Nam Lee, Seong-Won Jo, Gul Rahim, Sang-Gug Lee, and Kyeongha Kwon.
\newblock Small-perturbation electrochemical impedance spectroscopy system with
  high accuracy for high-capacity batteries in electric vehicles.
\newblock {\em IEEE Transactions on Industrial Electronics}, pages 1--11, 2024.

\bibitem{DandC}
D.J. Leith, W.E. Leithead, E.~Solak, and R.~Murray-Smith.
\newblock Divide \& conquer identification using gaussian process priors.
\newblock In {\em Proceedings of the 41st IEEE Conference on Decision and
  Control, 2002.}, volume~1, pages 624--629 vol.1, 2002.

\bibitem{LTVBFE}
Ebrahim Louarroudi, John Lataire, Rik Pintelon, Pieter Janssens, and Jan
  Swevers.
\newblock Frequency domain, parametric estimation of the evolution of the
  time-varying dynamics of periodically time-varying systems from noisy
  input--output observations.
\newblock {\em Mechanical Systems and Signal Processing}, 47(1-2):151--174,
  2014.

\bibitem{mangan2017model}
Niall~M Mangan, J~Nathan Kutz, Steven~L Brunton, and Joshua~L Proctor.
\newblock Model selection for dynamical systems via sparse regression and
  information criteria.
\newblock {\em Proceedings of the Royal Society A: Mathematical, Physical and
  Engineering Sciences}, 473(2204):20170009, 2017.

\bibitem{micchelli2005learning}
Charles~A Micchelli and Massimiliano Pontil.
\newblock On learning vector-valued functions.
\newblock {\em Neural computation}, 17(1):177--204, 2005.

\bibitem{mukherjee2006learning}
Sayan Mukherjee, Ding-Xuan Zhou, and John Shawe-Taylor.
\newblock Learning coordinate covariances via gradients.
\newblock {\em Journal of Machine Learning Research}, 7(3), 2006.

\bibitem{ndaoud2020optimal}
Mohamed Ndaoud and Alexandre~B Tsybakov.
\newblock Optimal variable selection and adaptive noisy compressed sensing.
\newblock {\em IEEE Transactions on Information Theory}, 66(4):2517--2532,
  2020.

\bibitem{olucha2024reduction}
E~Javier Olucha, Bogoljub Terzin, Amritam Das, and Roland T{\'o}th.
\newblock On the reduction of linear parameter-varying state-space models.
\newblock {\em arXiv preprint arXiv:2404.01871}, 2024.

\bibitem{piga2013lpv}
Dario Piga and Roland T{\'o}th.
\newblock Lpv model order selection in an ls-svm setting.
\newblock In {\em 52nd IEEE Conference on Decision and Control}, pages
  4128--4133. IEEE, 2013.

\bibitem{piironen2016projection}
Juho Piironen and Aki Vehtari.
\newblock Projection predictive model selection for gaussian processes.
\newblock In {\em 2016 IEEE 26th international workshop on machine learning for
  signal processing (MLSP)}, pages 1--6. IEEE, 2016.

\bibitem{RKHSgnr2}
Gianluigi Pillonetto, Francesco Dinuzzo, Tianshi Chen, Giuseppe De~Nicolao, and
  Lennart Ljung.
\newblock Kernel methods in system identification, machine learning and
  function estimation: A survey.
\newblock {\em Automatica}, 50(3):657--682, 2014.

\bibitem{PINTELON20112892}
R.~Pintelon, G.~Vandersteen, J.~Schoukens, and Y.~Rolain.
\newblock Improved (non-)parametric identification of dynamic systems excited
  by periodic signals—the multivariate case.
\newblock {\em Mechanical Systems and Signal Processing}, 25(8):2892--2922,
  2011.

\bibitem{SysIDFreqB}
Rik Pintelon and Johan Schoukens.
\newblock {\em System identification: a frequency domain approach}.
\newblock John Wiley \& Sons, 2012.

\bibitem{8194844}
Antonino Riccobono, Markus Mirz, and Antonello Monti.
\newblock Noninvasive online parametric identification of three-phase ac power
  impedances to assess the stability of grid-tied power electronic inverters in
  lv networks.
\newblock {\em IEEE Journal of Emerging and Selected Topics in Power
  Electronics}, 6(2):629--647, 2018.

\bibitem{rizvi2018model}
Syed~Z Rizvi, Farshid Abbasi, and Javad~Mohammadpour Velni.
\newblock Model reduction in linear parameter-varying models using autoencoder
  neural networks.
\newblock In {\em 2018 Annual American Control Conference (ACC)}, pages
  6415--6420. IEEE, 2018.

\bibitem{rizvi2016kernel}
Syed~Z Rizvi, Javad Mohammadpour, Roland T{\'o}th, and Nader Meskin.
\newblock A kernel-based pca approach to model reduction of linear
  parameter-varying systems.
\newblock {\em IEEE Transactions on Control Systems Technology},
  24(5):1883--1891, 2016.

\bibitem{rojas2014sparse}
Cristian~R Rojas, Roland T{\'o}th, and H{\aa}kan Hjalmarsson.
\newblock Sparse estimation of polynomial and rational dynamical models.
\newblock {\em IEEE Transactions on Automatic Control}, 59(11):2962--2977,
  2014.

\bibitem{rosasco2013nonparametric}
Lorenzo Rosasco, Silvia Villa, Sofia Mosci, Matteo Santoro, and Alessandro
  Verri.
\newblock Nonparametric sparsity and regularization.
\newblock {\em Journal of Machine Learning Research}, 14:1665--1714, 2013.

\bibitem{scholkopf2001generalized}
Bernhard Sch{\"o}lkopf, Ralf Herbrich, and Alex~J Smola.
\newblock A generalized representer theorem.
\newblock In {\em International conference on computational learning theory},
  pages 416--426. Springer, 2001.

\bibitem{schoukens2019nonlinear}
Johan Schoukens and Lennart Ljung.
\newblock Nonlinear system identification: A user-oriented road map.
\newblock {\em IEEE Control Systems Magazine}, 39(6):28--99, 2019.

\bibitem{schoukens2015structure}
Johan Schoukens, Rik Pintelon, Yves Rolain, Maarten Schoukens, Koen Tiels,
  Laurent Vanbeylen, Anne {Van Mulders}, and Gerd Vandersteen.
\newblock Structure discrimination in block-oriented models using linear
  approximations: A theoretic framework.
\newblock {\em Automatica}, 53:225--234, 2015.

\bibitem{JointPaper}
Mehrad~Ghasem Sharabiany, Sadegh Ebrahimkhani, and John Lataire.
\newblock Nonlinear continuous-time system identification by linearization
  around a time-varying setpoint.
\newblock {\em International Journal of Robust and Nonlinear Control},
  34(15):10454--10479, 2024.

\bibitem{singh2021learning}
Sumeet Singh, Spencer~M Richards, Vikas Sindhwani, Jean-Jacques~E Slotine, and
  Marco Pavone.
\newblock Learning stabilizable nonlinear dynamics with contraction-based
  regularization.
\newblock {\em The International Journal of Robotics Research},
  40(10-11):1123--1150, 2021.

\bibitem{sontag1999notions}
Eduardo~D Sontag and Yuan Wang.
\newblock Notions of input to output stability.
\newblock {\em Systems \& Control Letters}, 38(4-5):235--248, 1999.

\bibitem{Allen3}
Michael~W. Sracic and Matthew~S. Allen.
\newblock Identifying parameters of multi-degree-of-freedom nonlinear
  structural dynamic systems using linear time periodic approximations.
\newblock {\em Mechanical Systems and Signal Processing}, 46(2):325--343, 2014.

\bibitem{tibshirani1996regression}
Robert Tibshirani.
\newblock Regression shrinkage and selection via the lasso.
\newblock {\em Journal of the Royal Statistical Society Series B: Statistical
  Methodology}, 58(1):267--288, 1996.

\bibitem{toth2010modeling}
Roland T{\'o}th.
\newblock {\em Modeling and identification of linear parameter-varying
  systems}, volume 403.
\newblock Springer, 2010.

\bibitem{toth2009order}
Roland T{\'o}th, Christian Lyzell, Martin Enqvist, Peter~SC Heuberger, and
  Paul~MJ Van~den Hof.
\newblock Order and structural dependence selection of lpv-arx models using a
  nonnegative garrote approach.
\newblock In {\em Proceedings of the 48h IEEE Conference on Decision and
  Control (CDC) held jointly with 2009 28th Chinese Control Conference}, pages
  7406--7411. IEEE, 2009.

\bibitem{van2018revealing}
Mark van~de Ruit, Gaia Cavallo, John Lataire, Frans~Ct Van Der~Helm, Winfred
  Mugge, Jan-Willem van Wingerden, and Alfred~C Schouten.
\newblock Revealing time-varying joint impedance with kernel-based regression
  and nonparametric decomposition.
\newblock {\em IEEE Transactions on Control Systems Technology},
  28(1):224--237, 2018.

\bibitem{van2020data}
Henk~J Van~Waarde, Jaap Eising, Harry~L Trentelman, and M~Kanat Camlibel.
\newblock Data informativity: a new perspective on data-driven analysis and
  control.
\newblock {\em IEEE Transactions on Automatic Control}, 65(11):4753--4768,
  2020.

\bibitem{weston2003use}
Jason Weston, Andr{\'e} Elisseeff, Bernhard Sch{\"o}lkopf, and Mike Tipping.
\newblock Use of the zero norm with linear models and kernel methods.
\newblock {\em The Journal of Machine Learning Research}, 3:1439--1461, 2003.

\bibitem{westwick2018using}
David~T Westwick, Gabriel Hollander, Kiana Karami, and Johan Schoukens.
\newblock Using decoupling methods to reduce polynomial narx models.
\newblock {\em IFAC-PapersOnLine}, 51(15):796--801, 2018.

\bibitem{wittwar2018interpolation}
Dominik Wittwar, Gabriele Santin, and Bernard Haasdonk.
\newblock Interpolation with uncoupled separable matrix-valued kernels.
\newblock {\em arXiv preprint arXiv:1807.09111}, 2018.

\bibitem{yamada2014high}
Makoto Yamada, Wittawat Jitkrittum, Leonid Sigal, Eric~P Xing, and Masashi
  Sugiyama.
\newblock High-dimensional feature selection by feature-wise kernelized lasso.
\newblock {\em Neural computation}, 26(1):185--207, 2014.

\bibitem{zhou2021gene}
Fei Zhou, Jie Ren, Xi~Lu, Shuangge Ma, and Cen Wu.
\newblock Gene--environment interaction: A variable selection perspective.
\newblock {\em Epistasis: Methods and Protocols}, pages 191--223, 2021.

\bibitem{LTPfNL}
İsmail Uyanık, Mustafa~M. Ankaralı, Noah~J. Cowan, Uluç Saranlı, and Ömer
  Morgül.
\newblock Identification of a vertical hopping robot model via harmonic
  transfer functions.
\newblock {\em Transactions of the Institute of Measurement and Control},
  38(5):501--511, 2016.

\end{thebibliography}



\appendix
\section{LPV Model Order Selection Problem} \label{appx:LPVmdlordrOptprob}
The optimization problem to estimate sparse LPV coefficients is:
\begin{equation*} \label{eq:optimProbInputVar}
\begin{aligned}
\underset{\substack{\bm{\alpha}_i,\rho_j}}{\text{min}} &\left( \mathbf{e}^{H} \mathbf{W}^{-1} \mathbf{e} + \gamma_{\mathrm{reg}} \| \mathbf{c} \|^2_{\mathcal{H}_{\text{curl}}} + \gamma_{\mathrm{ord}} \sum_{j=1}^{n_x} \rho_j \right) \\
 &-\rho_j \leq  \hat{c}_j(\mathbf{p}_L(t'_s)) \leq \rho_j, \; \rho_j>0, \; \forall s \in \mathbb{N}_{n_t}
\end{aligned}
\end{equation*}
In this formulation, $t_k \in \mathbb{T}$, with $k \in \mathbb{N}_N$, represents a sample point; $t'_s \in \mathbb{T}'$, with $s \in \mathbb{N}_{n_t}$, denotes a nodal point; and $\check{t}_i \in \mathbb{T}_{\mathrm{ord}} = \mathbb{T} \cup \mathbb{T}'$ is the $i$-th element of $\mathbb{T}_{\mathrm{ord}}$. 

\textbf{(I) Data-fit term:}  
The vector $\mathbf{e}$ denotes the error of the LPV equation \eqref{eq:LPVeqErr}. In \eqref{eq:LPVcoeffAgebEq}, the unknown LPV coefficient vector $\mathbf{c}_{\mathrm{LPV}}$ is replaced by its representer $\hat{\mathbf{c}} \in \mathcal{V}_{\mathrm{ord}}$ (see Theorem~\ref{thm:LPVmodelOrderRepresenter}), which is evaluated at the sample points $t_k \in \mathbb{T}$:
 $\hat{\mathbf{c}}(\mathbf{p}_L(t_k)) =  \sum_{i=1}^{N_{\mathrm{ord}}}  \mathbf{K}_{\mathrm{curl}}(\mathbf{p}_L(t_k),\mathbf{p}_L(\check{t}_i)) \bm{\alpha}_i$.

\textbf{(II) Constraint:}  
Here, $\hat{c}_j$ denotes the $j$-th element of $\hat{\mathbf{c}}$ evaluated at the nodal points $t'_s \in \mathbb{T}'$: $\hat{c}_j(\mathbf{p}_L(t'_s)) = \sum_{i=1}^{N_{\mathrm{ord}}}  \mathbf{K}_{\mathrm{curl}}(\mathbf{p}_L(t'_s),\mathbf{p}_L(\check{t}_i))_{j,:} \bm{\alpha}_i$.

\textbf{(III) Regularization term:}  
Using the representer $\hat{\mathbf{c}}\in\mathcal{V}_{\mathrm{ord}}$, the regularization term becomes:
{\footnotesize
\begin{equation*}
\begin{aligned}
& \| \mathbf{c} \|^2_{\mathcal{H}_{\mathrm{curl}}} = \left\| \sum_{i=1}^{N_{\mathrm{ord}}}  \mathbf{K}_{\mathrm{curl}}(\cdot,\mathbf{p}_L(\check{t}_i)) \bm{\alpha}_i  \right\|^2_{\mathcal{H}_{\mathrm{curl}}} =\\
& \langle \sum_{i=1}^{N_{\mathrm{ord}}}  \mathbf{K}_{\mathrm{curl}}(\cdot,\mathbf{p}_L(\check{t}_i)) \bm{\alpha}_i  , \sum_{l=1}^{N_{\mathrm{ord}}}  \mathbf{K}_{\mathrm{curl}}(\cdot,\mathbf{p}_L(\check{t}_l)) \bm{\alpha}_l  \rangle_{\mathcal{H}_{\mathrm{curl}}} \\
&= \sum_{i=1}^{N_{\mathrm{ord}}} \sum_{l=1}^{N_{\mathrm{ord}}} \bm{\alpha}^{\top}_i \mathbf{K}_{\mathrm{curl}}(\mathbf{p}_L(\check{t}_i),\mathbf{p}_L(\check{t}_l)) \bm{\alpha}_l 
\end{aligned}
\end{equation*}
}
\section{LPV Scheduling Reduction Problem} \label{appx:LPVschreducOptprob}
The optimization problem for selecting the LPV scheduling dependencies is formulated as:
\begin{equation*} \label{eq:optimProbschedselect}
\begin{aligned}
\underset{\substack{\bm{\alpha}_i,\bm{\alpha}'_{j,k},\tau_{j,l}}}{\text{min}} &\left( \mathbf{e}^{H} \mathbf{W}^{-1} \mathbf{e} + \gamma_{\mathrm{reg}} \| \mathbf{c} \|^2_{\mathcal{H}_{\mathrm{curl}}} + \gamma_{\mathrm{sch}} \sum_{j=1}^{n_x} \sum_{l=j}^{n_x} \tau_{j,l} \right) \\
 &-\tau_{j,l} \leq  \frac{\partial   \hat{c}_j(\mathbf{p}'_L(t'_s))}{\partial p'_l} \leq \tau_{j,l}, \; \tau_{j,l}>0, \; \forall s \in \mathbb{N}_{n_t}
\end{aligned}
\end{equation*}
We begin by introducing:
$ \mathbf{p}'_L \coloneqq  \begin{bmatrix}
p'_{1} & p'_{2} & \dots & p'_{n_x}
\end{bmatrix} ^\top
=  
 \begin{bmatrix} y_L(t) & \dots & y_L^{(n_a)}(t) & u_L(t) & \dots & u_L^{(n_b)}(t) \end{bmatrix} ^{\top}
$, which provides an alternative notation for $\mathbf{p}_L$ in \eqref{eq:PLaspi}.

\textbf{(I) Data-Fit Term:}  
The vector $\mathbf{e}$ in \eqref{eq:LPVeqErr} represents the error in the LPV equation. In \eqref{eq:LPVcoeffAgebEq}, the unknown LPV coefficient vector $\mathbf{c}_{\mathrm{LPV}}$ is replaced by the representer $\hat{\mathbf{c}} \in \mathcal{V}_{\mathrm{sch}}$ (see Theorem~\ref{thm:LPVschvariableRepresenter}), which is evaluated at the data points $t_k \in \mathbb{T}$. This gives: 
$
 \hat{\mathbf{c}}(\mathbf{p}_L(t_k)) =\sum_{i=1}^{N} \mathbf{K}_{\mathrm{curl}}(\mathbf{p}_L(t_k), \mathbf{p}_L(t_i)) \bm{\alpha}_i +  
 \sum_{l=1}^{n_x} \sum_{k=1}^{n_t} \frac{\partial \mathbf{K}_{\mathrm{curl}}(\mathbf{p}_L(t_j), \mathbf{p}_L(t'_k))}{\partial p_l} \bm{\alpha}'_{l,k}
$.

\textbf{(II) Partial Derivative Constraints:}  
For the partial derivative constraints, the $j$th LPV coefficient computed from the representer in Theorem~\ref{thm:LPVschvariableRepresenter} and evaluated at the nodal points $t'_s \in \mathbb{T}'$ is:
 $
 \hat{c}_j(\mathbf{p}'_L(t'_s)) = \sum_{i=1}^{N}  \mathbf{K}_{\mathrm{curl}}(\mathbf{p}'_L(t'_s)),\mathbf{p}_L(t_i))_{j,:} \bm{\alpha}_i + 
  \sum_{l=1}^{n_x} \sum_{k=1}^{n_t} \frac{\partial   \mathbf{K}_{\mathrm{curl}}(\mathbf{p}'_L(t'_s)),\mathbf{p}_L(t'_k))_{j,:}}{\partial p_l}  \bm{\alpha}'_{l,k}
$. 

Then, the partial derivative of the $j$th LPV coefficient with respect to $p'_a$ and evaluated at  $t'_s \in \mathbb{T}'$ is given by:
\begin{equation*} \label{eq:apppdofLPVcoefschselect}
\begin{aligned}
& \frac{\partial \hat{c}_j(\mathbf{p}'_L(t'_s))}{\partial p'_a} = \sum_{i=1}^{N}   \frac{\partial \mathbf{K}_{\mathrm{curl}}(\mathbf{p}'_L(t'_s),\mathbf{p}_L(t_i))_{j,:} }{\partial p'_a}\bm{\alpha}_i  \\
& +  \sum_{l=1}^{n_x} \sum_{k=1}^{n_t} \frac{\partial^2  \mathbf{K}_{\mathrm{curl}}(\mathbf{p}'_L(t'_s),\mathbf{p}_L(t'_k))_{j,:}}{\partial p'_a \partial  p_l}  \bm{\alpha}'_{l,k}
\end{aligned}
\end{equation*}
\textbf{(III) Regularization Term:}  
Using the representer in Theorem~\ref{thm:LPVschvariableRepresenter}, the regularization term is expressed as:
{\footnotesize
\begin{equation*}
\begin{aligned}
& \| \mathbf{c} \|^2_{\mathcal{H}_{\mathrm{curl}}} = \| \sum_{i=1}^{N}  \mathbf{K}_{\mathrm{curl}}(\cdot,\mathbf{p}_L(t_i)) \bm{\alpha}_i \\
& + \sum_{l=1}^{n_x}  \sum_{k=1}^{n_t}  \frac{\partial   \mathbf{K}_{\mathrm{curl}}(\cdot,\mathbf{p}_L(t'_k))}{\partial p_l}  \bm{\alpha}'_{l,k}   \|^2_{\mathcal{H}_{\mathrm{curl}}} =\\
& \langle \sum_{i=1}^{N}  \mathbf{K}_{\mathrm{curl}}(\cdot,\mathbf{p}_L(t_i)) \bm{\alpha}_i  , \sum_{k=1}^{N}  \mathbf{K}_{\mathrm{curl}}(\cdot,\mathbf{p}_L(t_k)) \bm{\alpha}_k  \rangle_{\mathcal{H}_{\mathrm{curl}}} + \\
&  2 \langle \sum_{i=1}^{N}  \mathbf{K}_{\mathrm{curl}}(\cdot,\mathbf{p}_L(t_i)) \bm{\alpha}_i ,  \sum_{l=1}^{n_x}  \sum_{k=1}^{n_t}  \frac{\partial   \mathbf{K}_{\mathrm{curl}}(\cdot,\mathbf{p}_L(t'_k))}{\partial p_l}  \bm{\alpha}'_{l,k}  \rangle_{\mathcal{H}_{\mathrm{curl}}} +\\
& \langle \sum_{l=1}^{n_x}  \sum_{k=1}^{n_t}  \frac{\partial   \mathbf{K}_{\mathrm{curl}}(\cdot,\mathbf{p}_L(t'_k))}{\partial p_l}  \bm{\alpha}'_{l,k},\sum_{i=1}^{n_x}  \sum_{s=1}^{n_t}  \frac{\partial   \mathbf{K}_{\mathrm{curl}}(\cdot,\mathbf{p}_L(t'_s))}{\partial p_i}  \bm{\alpha}'_{i,s} \rangle_{\mathcal{H}_{\mathrm{curl}}}\\
&= \sum_{i=1}^{N} \sum_{k=1}^{N} \bm{\alpha}^{\top}_i \mathbf{K}_{\mathrm{curl}}(\mathbf{p}_L(t_i),\mathbf{p}_L(t_k)) \bm{\alpha}_k  + \\
& 2 \sum_{l=1}^{n_x} \sum_{i=1}^{N} \sum_{k=1}^{n_t}  \bm{\alpha}_i^{\top}  \frac{\partial   \mathbf{K}_{\mathrm{curl}}(\mathbf{p}'_L(t_i),\mathbf{p}_L(t'_k))}{\partial p_l}  \bm{\alpha}'_{l,k} + \\
& \sum_{l=1}^{n_x} \sum_{k=1}^{n_t} \sum_{i=1}^{n_x}  \sum_{s=1}^{n_t} \bm{\alpha}_{i,k} ^{'\top}  \frac{\partial^2   \mathbf{K}_{\mathrm{curl}}(\mathbf{p}'_L(t'_k),\mathbf{p}_L(t'_s))}{\partial p'_l \partial p_i} \bm{\alpha}'_{i,s} 
\end{aligned}
\end{equation*}
}

\end{document}